\documentclass[useAMS,usenatbib]{mn2e}
\usepackage{graphicx}
\usepackage{natbib}
\usepackage{epsfig}
\newcommand{\bmth}[1]{\mbox{\boldmath${#1}$}}

\title[Tidal capture through perturbation of rotating stars ]
{Dynamical tides excited in  rotating stars of different masses and ages and the formation of close in  orbits}
\author[S.V. Chernov,  J. C. B. Papaloizou, P. B. Ivanov]{S. V. Chernov
$^{1}$\thanks{E-mail: chernov@td.lpi.ru (SVCh)}, J. C. B. Papaloizou $^{2}$\thanks{E-mail:
J.C.B.Papaloizou@damtp.cam.ac.uk (JCBP)}  and  P.B.Ivanov$^{1}$\thanks{E-mail:
pbi20@cam.ac.uk (PBI)} \\
$^{1}$Astro Space Centre, P.N. Lebedev Physical Institute, 84/32
Profsoyuznaya Street, Moscow, 117997, Russia  \\
$^{2}$ DAMTP, Centre for Mathematical Sciences, University of
Cambridge, Wilberforce Road, Cambridge CB3 0WA }

\begin{document}

\date{Accepted. Received; in original form}

\pagerange{\pageref{firstpage}--\pageref{lastpage}} \pubyear{2010}

\maketitle

\label{firstpage}

\begin{abstract}
We study  the tidal response
of rotating solar mass stars, as well as  more massive rotating stars, of different ages in the context
of tidal captures leading to either  giant exoplanets on close in orbits, or the formation of binary systems
in star clusters. To do this, we adopt approaches based on normal mode and associated overlap integral
evaluation, developed in a companion paper by Ivanov et al., and  direct numerical simulation,  to evaluate
energy and angular momentum exchanges between the orbit and normal modes.
The two approaches are found to be in essential agreement apart from when encounters occur
near to pseudosynchronization, where the stellar angular velocity  and the orbital angular velocity at periastron are
approximately matched. We find that the strength of tidal interaction being expressed in dimensionless natural units
is significantly weaker for the more massive stars, as compared
to the  solar mass stars, because of the
lack of significant convective envelopes in the former case. On the other hand the interaction is found to be stronger
for retrograde as opposed to prograde orbits in all cases. In addition, for a given pericentre distance,
 tidal interactions also strengthen for more evolved
stars on account of their radial expansion. In agreement with previous work
based on simplified polytropic models,  we find that energy transferred to their central stars could play  a significant
role in the early stages of the circularisation of potential 'Hot Jupiters'.

\end{abstract}


\section{Introduction}
Tidal interaction
leads to synchronisation and
orbital circularisation  of close binary stars (eg. Zahn 1977, Hut 1981).
It may also result in double star or star-planet systems
that undergo close encounters in   marginally unbound orbits
becoming tidally captured into highly eccentric orbits that then begin to circularise
(eg. Press \& Teukolsky 1977).  A process of this kind is believed to account for giant exoplanets
in close orbits with periods of a few days
 (eg. Weidenschilling \&  Marzari 1996,
Rasio \& Ford 1996).

The determination of  the tidal evolution
requires the calculation of the  response of the tidally perturbed body.
This involves energy and angular momentum exchange
between its  normal modes and the orbit leading to its  evolution.
In a companion paper, Ivanov et al. (2013),  subsequently referred to as Paper 1, we developed general procedures for calculating tidal
energy and angular momentum exchange rates, for bodies in periodic orbits,
that  are  associated with  an identifiable
regular spectrum of low frequency  rotationally
modified gravity modes, for rotating stars with realistic
structure.

These are likely to give rise to the dominant tidal
response in bodies with stratification, where the tidal forcing
frequencies significantly exceed the inverse of the convective
time scale associated with any convection zone, so that any
effective turbulent viscosity  is inefficient.
This is also expected to be the case  for rotating stars,
when the dominant tidal forcing frequencies as viewed in the rotating frame
exceed twice the rotation frequency with the consequence that inertial modes are not efficiently
excited in convective regions.

In Paper 1 we  also gave  expressions from which the energy and
angular momentum transferred to stellar modes of oscillation as a result of
parabolic
encounters can be calculated. A process that could lead to tidal captures
and also governs the initial phase of orbital circularisation when the
orbit is very eccentric (eg. Ivanov \& Papaloizou 2004).
Evaluation of  the response arising from  normal modes requires calculation
of  mode eigenfrequencies and corresponding overlap integrals that determine the strength of  mode
coupling with the tidal potential
(eg.  Press \&
Teukolsky 1977). This procedure was  discussed in some detail in Paper 1  for the case when the
traditional approximation, appropriate for low frequency  modes in stratified layers, was adopted.
We remark that, as  discussed in more detail in Paper 1, tidal phenomena such as  energy and angular momentum exchange
through  parabolic encounters, or orbital
evolution in the regime of so-called moderately strong
viscosity (eg. Zahn 1977, Goodman $\&$ Dickson 1998), where propagating rotationally modified gravity waves
attain short wavelengths, and so are dissipated before reaching  boundaries from which they can be reflected,
are such that results  are independent  of the  precise specification of the dissipation process.
In this regime, the wave  dissipation should also occur  on a time scale that is  significantly longer than the  locally  excited  wave period which will also be characteristic of the time for excitation due to tidal perturbation.

In this  paper we apply the formalism developed in Paper 1, where only Sun-like stars were considered,
to calculate the normal modes and their associated
overlap integrals for a range of tidal forcing frequencies for two
models of a rotating solar mass star with  different ages, as well as several models of more massive rotating stars, with  different ages.
The dependence of these quantities on the existence  and extent of convective regions
and the transition between convective and radiative regions is elucidated.
We also compare results obtained from the normal mode approach of Paper 1
to those obtained from direct numerical simulations of parabolic encounters (eg. Papaloizou \& Ivanov 2010)
delineating when there is good agreement between the two approaches.
Our  results are then applied to the tidal capture and initial orbital circularisation
of giant exoplanets for both prograde and retrograde orbits
and also  the  tidal capture of stars to form binary systems in stellar clusters
(eg. Fabian, Pringle $\&$ Rees 1975,  Press $\&$ Teukolsky 1977).

The plan of the paper is as follows.
In section 2 we describe the stellar models for which we calculated the quantities
that enable their  exchange of energy and angular momentum under
tidal  gravitational  perturbation  due to a companion to be calculated.
These quantities are the overlap integrals and the low frequency rotationally modified
$g$ mode spectrum and they are discussed in detail in Paper 1.
We consider models in the range of $1-5M_{\odot}$ with a variety of ages.
As indicated in Paper 1, the extent of any convective envelope and/or core plays a significant
role in determining the strength of tidal interaction as  also does the detailed form of the
transition between convective and radiative regions.

In section 3 we discuss the properties of the numerically calculated mode
spectra and overlap integrals for the stellar models considered. We also derive the rotational
splitting coefficients which give the first order shifts of mode eigenfrequencies as a result
of stellar rotation.  In the non rotating case, the overlap integrals were found to be markedly larger
for Sun-like stars as compared to either a polytrope with index 3 or more massive models
with much less extensive convective regions. This is because of the convective envelope
and is expected from the theory developed in Paper 1.

The overlap integrals are also calculated for rotating models under the neglect of centrifugal distortion
and the adoption of the traditional approximation as indicated in Paper 1. Results for  angular velocities
of rotation in  units of the critical rotation rate in  the range  $0.1-0.4$ are presented.

We go on to apply our results to evaluate the energy and angular momentum
exchanged as a result of a parabolic encounter with a companion. These enable the  possibility of tidal capture
from unbound orbits to be assessed. In addition the time scale for the initial stages of orbital
circularisation to occur for low  planetary mass companions  is  estimated.

We  compare energy and angular momentum transfers obtained  through the normal
mode/overlap integral approach to results obtained from solving the encounter problem as an initial
value problem  numerically (Papaloizou \& Ivanov 2010;  Ivanov \& Papaloizou 2011)
 for the full  range of rotation rates and  for pericentre distances
that are not too large to make calculation intractable. Both prograde and retrograde encounters
are considered. We found that the methods are  in good agreement
apart from the situation where the system is close to pseudosynchronization.
In this case the effective tidal forcing frequencies are comparable to the rotation frequency
and inertial modes, not taken into account in the normal mode approach  can  play a significant role.
As the characteristic  tidal forcing frequencies are expected to be significantly
larger than stellar  rotation frequencies,  inertial modes are unlikely to be excited in the star during the
initial stages of the  formation of close in
giant planet orbits of 'Hot Jupiters'.  Accordingly we do not  pursue the issue of inertial modes
further in this paper.

Finally in section 5 we summarize and discuss our results.
We remark  that as in our earlier work (Ivanov \& Papaloizou 2011),  which considered polytropic models
with index, 3, we find that tidal interaction with the central star
is significantly stronger for retrograde orbits and that it
could play a significant
role in the circularisation process for giant planets,  so potentially reducing the amount
of potentially destructive energy dissipation in the planetary interior.

\section{Formulation of the problem and details of numerical methods}\label{Basiceq}

\subsection{Coordinate system and notation}

The basic definitions  and notation adopted in this paper are the same as in  Paper 1.
We use  either a spherical coordinate system $(r, \phi, \theta)$ or associated
cylindrical polar coordinate system $(\varpi, \phi, z)$
with origin at the centre of mass of the star.   When viewed in an inertial frame,
the unperturbed star   rotates uniformly  about the $z$ axis with angular velocity $\Omega$.
For our  reference frame,  we adopt  the rotating frame in which the unperturbed star appears at rest.

\subsection{Stellar models considered}\label{Modeldesc}
Our calculations are for rotating stars of different masses and  ages.  As in  paper 1
centrifugal distortion  is neglected  with the consequence that  equilibrium
structures are  not  modified by rotation.  Therefore
standard spherically symmetric  models are used.

\begin{table}
 \centering
 \begin{minipage}{100mm}
  \begin{tabular}{@{}llllr@{}}
  \hline
   Model  & Mass & Radius & Age & Mean density  \\
 \hline
 1p          &  1         & 1         &  & 1  \\
 \hline
 1a          &  1         & 0.91         & $1.67 \cdot 10^8$ &   1.33\\
 \hline
 1b         &  1          & 1   & $4.41\cdot 10^9$ & 1  \\
 \hline
 1.5a     &  1.5       & 2.08    & $1.27 \cdot 10^7$ &  0.166 \\
 \hline
 1.5b     &  1.5       & 1.46   & $5.96\cdot  10^7$ &   0.482 \\
 \hline
 1.5c     &  1.5       & 1.82    & $1.58\cdot  10^9$ & 0.249 \\
 \hline
 2a        &  2          & 2.68   & $6.81\cdot 10^6$ &  0.104 \\
 \hline
 2b        &  2          & 1.63   & $2.93\cdot 10^7$ & 0.462 \\
 \hline
 2c        &  2          & 2.25   & $5.93\cdot 10^8$ & 0.175 \\
 \hline
 2d       &  2          & 2.91   & $8.44 \cdot 10^8$ &  0.0811\\
 \hline
 5a       &  5          & 2.69   & $2.54 \cdot 10^6$ &  0.257 \\
 \hline
\end{tabular}
\end{minipage}
\caption{Radii, masses, ages and mean densities for models used
in our calculations. Radii, masses and mean densities are
expressed in units  corresponding to the present day
Sun, stellar ages are expressed in years.\label{t1}}
\end{table}

We consider  models of stars of masses  $M_*=1,$  $1.5,$ $2,$
and $5M_{\odot}$ of different  ages.  Stellar masses
are  expressed in  solar masses
($1M_{\odot}=1.9891\cdot 10^{33}g$).   Radii, $R_*,$  are expressed in  solar radii
($1R_{\odot}=6.9551\cdot 10^{10}cm$) and ages, expressed in years, are given in Table
\ref{t1}.  Additionally, we have calculated all quantities of
interest for a stellar model consisting of a   polytrope with index $n=3.$
 The mass and radius are scaled to solar values and the adiabatic index
$\Gamma=5/3.$ This  serves as
reference  model for our analysis and is referred  to as  model 1p.

 All realistic stellar models
apart from  model 1b were kindly provided to us by I.W.  Roxburgh.
The numerical code
used to obtain these models is
discussed in  Roxburgh (2008). Model 1b  is for  the present day
Sun. It is discussed in Christensen-Dalsgaard et al. (1996).
 Unlike models described elsewhere
in a similar context (see McMillan, McDermott $\&$ Taam 1987)
our models have metallicity appropriate for population I stars.
 The  zero age hydrogen
mass fraction X=0.7 and the mass fraction of heavy elements Z=0.02 for all models.

 Convective heat transport is  described by a standard form of mixing length theory
( Kippenhahn et al. 2013).    Mixing in convective and semi-convective zones
is dealt with by incorporating diffusion into the equations governing the evolution
of the  chemical abundances ( see Eggleton 1972).  The diffusion coefficient
is taken to be $D_{conv}= v_{conv}{l}/6,$ where $v_{conv}$ is the convective velocity
and ${ l}$ is the mixing length.

\begin{figure}
\begin{center}
\vspace{8cm}\includegraphics{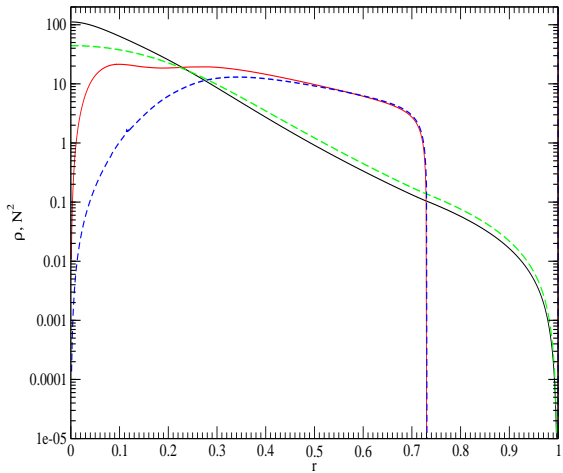}
\end{center}
\vspace{0.5cm} \caption{The density $\rho $ (in units of the mean
density $\tilde \rho = 3M_*/ (4\pi R_*^3)$) and square of the Brunt - V$\ddot
{\rm a}$is$\ddot {\rm a}$l$\ddot {\rm a}$ frequency,  $N^2$ (in
units of $GM_*/ R_*^3$) as functions of the radius $r$ expressed
in units of $R_*.$  The solid curves correspond to  model 1b
while the dashed curves are for the model 1a. The curves
monotonically decreasing with $r$ are for the density
distributions,  while the curves having maxima at some values of
$r$ are for the Brunt - V$\ddot {\rm a}$is$\ddot {\rm a}$l$\ddot
{\rm a}$ frequencies.} \label{Fig1}
\end{figure}
\begin{figure}
\begin{center}
\vspace{8cm}\includegraphics{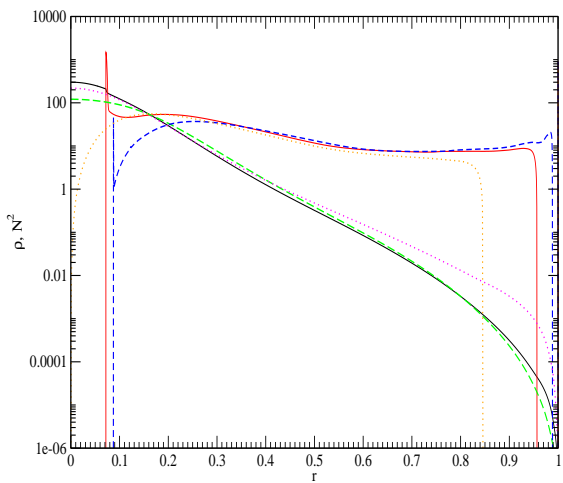}
\end{center}
\vspace{0.5cm} \caption{Same as in Fig. \ref{Fig1} but for models
with $M_*=1.5M_{\odot}$. Solid, dashed and dotted curves are for
models 1.5c, 1.5b and 1.5a, respectively. Note that there  are regions  very close to the surface
where  a
weak density inversion occurs  in these models.  However, the values of the density
 where this occurs are below  the minimum level plotted.} \label{Fig2}
\end{figure}
\begin{figure}
\begin{center}
\vspace{8cm}\includegraphics{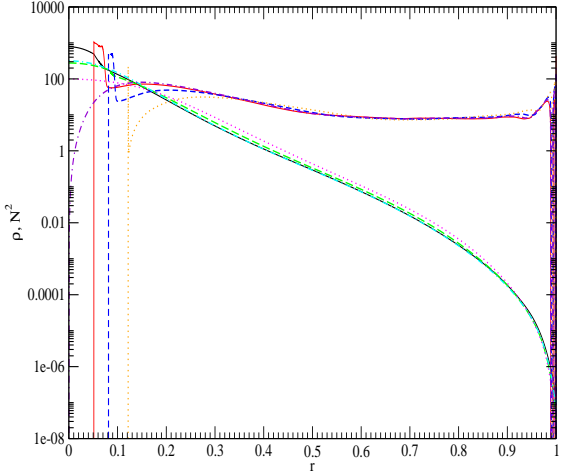}
\end{center}
\vspace{0.5cm} \caption{Same as in Fig. \ref{Fig1} but for models with $M_*=2M_{\odot}$. Solid, dashed, dotted and
dot-dashed curves
are for models 2d, 2c, 2b and 2a, respectively. Note that the dashed and dot-dashed curves for the density almost 
coincide} \label{Fig3}
\end{figure}
\begin{figure}
\begin{center}
\vspace{8cm}\includegraphics{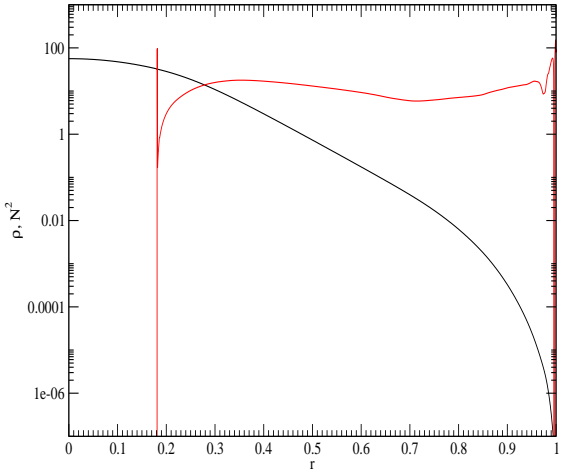}
\end{center}
\vspace{0.5cm} \caption{Same as in Fig. \ref{Fig1} but for model 5a with $M_*=5M_{\odot}$.} \label{Fig4}
\end{figure}

As is discussed in Paper 1 when low frequency gravity modes are
considered,  their important  properties are found to be
mainly determined by the  functional forms of the  density,
$\rho,$ and the Brunt - V$\ddot {\rm a}$is$\ddot {\rm a}$l$\ddot {\rm
a}$ frequency $N$. In particular,
the locations  of stably stratified radiative regions, where
$N^2 > 0,$ and the behaviour of $N$ in the neighbourhood  of
transitions  from stably stratified  to convective regions, are  particularly significant.
 For the models described here the functional forms in these transition regions  are strongly affected  by the evolutionary history
of the chemical composition profiles.  Although chemical diffusion is included, the coefficient
is zero in radiative regions and large in convective regions so that mixing is efficient there.
This results in the possibility of very rapidly varying or even discontinuous abundance profiles
according to how the boundary between a radiative  and convective zone moves with evolution.
Suppose this interface moves with speed $v_{I},$ we can form the dimensionless quantity
$v_{I}l/D_{conv}= 6v_{I}/v_{conv}\sim t_{conv}/t_{ev},$ the latter ratio being
the ratio of the convection time scale to the evolutionary  time scale.

On dimensionless grounds we might expect the width of the transition region, $w_{rc},$  between a convective and radiative zone
would be given by an expression of the form $ w_{rc} \sim lF(  t_{conv}/t_{ev}),$ where the form of the function $F$
is determined by the way the diffusion coefficient vanishes as the radiative zone is entered and the evolutionary history.
Simple modelling suggests that $F$ is a monotonic function of its argument, with the variation being more rapid for diffusion
coefficients with more rapid cutoffs.  For example a linear variation is expected  when the diffusion coefficient
 inside the radiative region  vanishes
as  $(r-r_{I})^2/l^2$, with $r_I$ being a distance of one mixing length away from a retreating convection zone boundary.
Note that, $t_{conv}/t_{ev} \sim 10^{-(9-10)}$, is very small leading to the possibility of very thin transition layers
when the evolutionary history is such that the composition in a convective zone differs significantly from its immediate surroundings.
At such a  transition  the mean molecular weight changes while hydrostatic equilibrium enforces continuity of the pressure.
There is a then a rapid change in the density with an associated spike in the density gradient and
the square of the Brunt-V$\ddot {\rm a}$is$\ddot {\rm a}$l$\ddot {\rm a}$ frequency.
  As  this  results from the density gradient,
 a large effect can be produced with there being only small jumps in the density or chemical profile.

Plots of  the density $\rho $ and the square of the Brunt -
V$\ddot {\rm a}$is$\ddot {\rm a}$l$\ddot {\rm a}$ frequency,
$N^2,$  against radius,  for the models used in our calculations
are given in Figs. \ref{Fig1}-\ref{Fig4}. In Fig. \ref{Fig1} we illustrate these two quantities
the two models with $M_*= 1M_{\odot}$ discussed in detail and adopted
in Paper 1. Note that negative values of $N^2$ that occur in convective regions
are not shown.
These  models exhibit a similar and
rather simple structure.  They have a radiative core
 and  convective envelope with a transition at a similar total radius fraction.
 The young Sun-like star,  model 1a,
  has a smaller radius than model 1b,
which corresponds to the present day Sun.  It  accordingly has a larger mean density
and is  therefore, less susceptible to tidal influence from
 a perturbing companion with a given orbital period.

Models 1.5a-1.5c, illustrated  in Fig. \ref{Fig2}, are for a star of mass
$M_*=1.5M_{\odot}$ at different ages. Their structure is more
complex  than that of Sun-like stars. The youngest  model 1.5a
(plotted with dotted curves) is similar to  models 1a and 1b in  that
it also has convective envelope with a convective core being absent.
 We remark that a weak density inversion occurs very close to the surface
in these models. This is a well known effect that occurs as hydrogen is ionised  in convective envelopes
 where the energy transport due to convection is inefficient (see eg. Latour 1970).
The ionisation zone is very thin in such cases,  and such that the reduction in the mean molecular weight,
that occurs with weak pressure variation there,  causes  the density inversion.
The density in these layers is very small and so they do not affect any of the tidal response calculations presented
in this paper significantly.
However, on account of the general similarity of the models,
we expect that the overlap integrals, $Q,$ characterising the
strength of tidal interactions and  discussed in detail in Paper 1,
will have similar properties to those obtained for  Sun-like stars.
But, the density at the base of convection zone,  expressed in terms
of the mean  density $\tilde \rho = 3M_*/ (4\pi R_*^3)$, $\rho_{cb},$ is smaller than
the corresponding quantity for  models 1a and 1b.  We find $\rho_{cb}\sim 10^{-3}\tilde \rho $
and $\sim 10^{-2}\tilde \rho$ for model 1.5a,  and  either of models 1a or  1b  respectively.

The analytic  WKBJ theory developed in Paper 1 predicts
that the overlap integrals, being  determined by the presence of a
convective envelope,  are proportional to $\sqrt \rho_{cb}$ in the
limit of sufficiently small eigenfrequencies (see equations (111),
(113) and (115) of Paper 1). Accordingly, we expect that
for a given value of the  eigenfrequency,
$Q$ values  for
 model 1.5a will be  smaller than those for  younger models,  by a factor
corresponding to  the  square root of the ratio of the respective values of $\rho_{cb}.$
 Here we recall that as the mode spectrum
is dense the overlap integrals are regarded as continuous functions of frequency.
 Let us note that  energy and angular momentum exchanges   due to tides
associated with a normal mode
are, in general,  proportional to the square of the appropriate overlap integral (see Paper 1). That means
that in a situation  where the contribution from the convective envelope  and the region of
its boundary with the inner radiative zone  determines the value of the overlap integral,   energy and angular momentum exchanges
 are approximately proportional to $\rho_{cb}$, see also  equation
(13) of  Goodman $\&$ Dickson (1998).

Our expectation is confirmed by
calculation, see Fig. \ref{Fig6}.
The  evolved models
1.5b and 1.5c have both convective envelopes and convective cores.
However, $\rho_c$ at the base of convective envelope is  relatively
small, being $\sim 4.6\cdot 10^{-5}$ for model
1.5c and $\sim 6\cdot 10^{-6}$ for model 1.5b. This  has the consequence that the
contribution to the overlap integrals associated with the
presence of a convective envelope  is strongly suppressed.

 Models 1.5b and 1.5c also have convective cores. But, the
transition region between the  radiative region and  the  convective core is
extremely sharp in these models  ( see discussion in Section \ref{Modeldesc} above), and must be considered as a
discontinuity for eigenfunctions with characteristic  wavelength larger
than  the typical size of the transition zone. The width of the latter, $\Delta r,$  is of the
order of or smaller than the grid size with  $\Delta r/r <
2.5\times  10^{-4}$, $2.5\times 10^{-5}$ for models 1.5b and 1.5c,
respectively. We emphasise that the detailed form of this transition region should be determined
from a complete treatment of convection,  including overshoot, see eg. Roxburgh (1978), Zahn (1991),
and  the effects of stellar rotation, etc. This cannot be undertaken
 at present\footnote{Note that   future advances  in astroseismology may lead to
some observational constraints on details of transitions between radiative and convective  regions
in the  near future, see eg. Silva Aguirre et al. (2011).}.
However, when eigenfunctions have  typical wavelengths larger
than the size of the transition zone, one cannot  assume that the
Brunt-V$\ddot {\rm a}$is$\ddot {\rm a}$l$\ddot {\rm a}$
frequency increases as a power of distance from the  boundary of the convective
region and perform a standard WKBJ analysis that assumes the  response  wavelength  is significantly  shorter than the  transition width.
 Accordingly  estimates of  quantities characterising
tidal interactions, such as  overlap integrals or related quantities, for
example the quantity $E_2$ used by Zahn (1977), based on this
assumption are not valid.

Let us estimate typical periods of modes, where  such  calculations are potentially  formally  invalid.
For definiteness we consider  model 1.5b.
We assume that  the  width of the  transition zone   is  as small  as suggested by our numerical model.
 From  Fig. \ref{Fig2} we see that for this model,
the value of Brunt - V$\ddot {\rm a}$is$\ddot {\rm a}$l$\ddot {\rm a}$
frequency at the maximum of the 'spike' close to the convective core  is $\approx 6.8$ and the radius of convective core
$r_c \approx 0.087$  in our dimensionless units.

From the WKBJ theory applied to  gravity waves,  the characteristic   wavelength, $\lambda $,  can be estimated  through
$\lambda \approx  r \omega /( \sqrt{6} N)$, where $\omega $ is the eigenfrequency and it is assumed that the star
is non-rotating, see eg. Christensen-Dalsgaard
(1998). On the other hand $\lambda  $ should be larger than the width of the transition region, estimated above
as  $ \sim 2.5\times  10^{-4}r$ or smaller. Thus we obtain
$\omega > 4.2\cdot 10^{-3}$ in  natural units for this inequality to be valid.
From table 1 it follows that this corresponds to periods $ < 40 \, days$. This means that  estimates based on a standard WKBJ analysis
 may be inapplicable  for all periods of interest.  Let us stress, however, that the width of the transition
region may be much larger than was assumed in order to obtain this estimate (see discussion in Section \ref{Modeldesc}).

In Fig. \ref{Fig3} we show the forms of the density and the
square of Brunt - V$\ddot {\rm a}$is$\ddot {\rm
a}$l$\ddot {\rm a}$ frequency for models with $M_*=2M_{\odot}$.
The youngest model, model 2a (dot-dashed curve) is fully radiative.
We expect that for this model, overlap integrals are
suppressed at low  frequencies as compared to models with
convective regions.
The  overlap integrals associated with such models  are expected to  decrease with
eigenfrequency, $\omega $, faster than any power of $\omega,$
as happens for a polytropic star represented as model 1p.
Thus, tidal interactions determined by low frequency gravity modes
become rather inefficient in this case. Models 2b-2d are similar to
those with mass $M_*=1.5M_{\odot}$ with the
difference  that the convective envelope is practically absent.
 Therefore, contributions to the overlap
integrals coming from the envelope  region should be very small. There are
also  almost  discontinuous transitions from radiative
envelopes  to convective cores. Thus we  conclude that previous estimates
of the strength of tidal interactions based on the regular
behaviour of $N^2$ close to this transition may  need revision.

Finally, in Fig \ref{Fig4} we plot the density and square of the
 Brunt - V$\ddot {\rm a}$is$\ddot {\rm
a}$l$\ddot {\rm a}$ frequency for a model of a young star with
$M_*=5M_{\odot}$. The structure of this model is rather simple and
similar to the cases of evolved stars with $M_*=2M_{\odot}$. There
is no convective envelope in this model and, again, there is a
quite sharp transition between the radiative region and convective
core.

\section{Properties of stellar eigenmodes: eigenspectra, overlap
integrals and rotational splitting coefficients}

In this Section we consider  the quantities  that
 determine the tidal interactions for given
orbital parameters and properties of decay of free stellar oscillation
either due to viscosity or non-linear effects.
These are the
eigenfrequencies of free pulsations $\omega$, normalised overlap
integrals $\hat Q$, and, in case of small rotational frequency
$\Omega \ll \Omega_*\equiv \sqrt{{GM_*/ R_*^3}}$,  the coefficients
$\beta$ which  determine the splitting of eigenfrequencies due to rotation
in the non-rotating frame.  We discus how these quantities    depend on stellar
structure

 The overlap integrals are discussed in detail in Paper 1. Here we briefly recall
them for completeness. In general,  $\hat Q$ is given by the expression
\begin{equation}
\hat Q=Q/\sqrt{n}, \quad Q=\left({\bmth\xi}{ |}\int_{0}^{2\pi} d\phi e^{-im\phi}\nabla
(r^{2}Y^{m}_{2})\right),
\label{q1}
\end{equation}
where it is implied that the inner scalar product of any two complex
vectors ${\bmth{ \eta}_{1}}$ and ${\bmth{ \eta}_{2}}$ is determined by integration over
the cylindrical coordinates $\varpi $ and $z$ as
 \begin{equation}
({\bmth{ \eta}}_{1}|{\bmth{ \eta}}_{2})=\int \varpi d\varpi dz\rho
({\bmth{ \eta}}_{1}^{*}\cdot {\bmth{ \eta}}_{2}), \label{q2}
\end{equation}
Here $*$ denotes  the complex conjugate, $Y^m_2$ is the spherical function, ${\bmth{ \xi}}$
is  the Lagrangian displacement vector corresponding to a particular eigenmode with eigenfrequency
$\omega$. It is assumed that in all expressions the dependence of ${\bmth{ \xi}}$ on
the azimuthal angle $\phi$, ${\bmth{ \xi}} \propto e^{im\phi}$, is factored out.
In general the azimuthal mode number, $m,$ is such that $|m| \leq 2.$ However,
we shall consider only $|m|=2$ below as this is the most important case (see Paper 1).

The norm $n$ is
determined by the expression
\begin{equation}
n=\pi (({\bmth {\xi}}| {\bmth
{\xi}})+({\bmth {\xi}}|{\bmth {\cal
C}}{\bmth{\xi}})/\omega^2), \label{q3}
\end{equation}
where ${\bmth {\cal C}}$ is an integro-differential self-adjoint operator, which when operating
on $-\bmth{\xi}$  gives the  restoring  acceleration due to the action
of gravity and pressure forces.

When the star is non-rotating the overlap integrals reduce  to the form given by
Press $\&$ Teukolsky (1977). A similar expression can be obtained in the so-called traditional
approximation discussed below and in Paper 1. In that case we have
\begin{equation}
Q=\alpha Q_{r}, {\rm with}  \quad Q_{r}=\int^{R_*}_{0}r^3dr \rho (2\xi
+\Lambda \xi^S), \label{q4}
\end{equation}
and
$n=n_{st}+n_r, $
\begin{equation}
{\rm  where } \hspace{1mm}  n_{st}=\int_0^{R_*} r^2dr \rho (\xi_j^2+\Lambda
(\xi^S)^2), {\rm and}\hspace{1mm} n_r=\nu I_r I_{\theta} \label{q5}
\end{equation}
with $I_r = \int_0^{R_*} r^2dr \rho (\xi^S)^2$ and $\nu=2\Omega/\omega$. In the expressions
(\ref{q4}) and (\ref{q5}) $\xi (r) $ and $\xi^S(r) $ give  the radial dependences  of the radial  component of the displacement vector
and the  angular   components,  respectively,   in the traditional
approximation.  The quantities $\Lambda $, $\alpha $ and $I_{\theta}$ are functions of $\nu $,
their explicit form and the dependence on $\nu $ are discussed in Paper 1.  Note that $\Lambda (\nu )$ is
an eigenvalue   associated with acceptable solutions of the  Laplace tidal equation.

When the star  does not
rotate,  the Laplace tidal equation
reduces to the Legendre equation,
the  solution of which  is  the  associated Legendre function.  We  then have $\alpha=1$,  $\Lambda=6$, $n=n_{st}$ and
the expressions (\ref{q4}) and (\ref{q5})  reduce to
their standard form as given by  e.g. Press $\&$ Teukolsky (1977) and Ivanov $\&$ Papaloizou (2004).

Hereafter we express all quantities of
interest in natural units. Thus, eigenfrequencies are expressed in terms of
the  natural stellar frequency $\Omega_*,$ and the overlap integrals in
terms of $\sqrt{M_*}R_*.$

\subsection{Eigenfrequencies and overlap integrals for non-rotating stars}

Let us first discuss the eigenfrequencies and overlap integrals for
non-rotating models. In this case  we set
$\alpha_i=1$ and $n_r=0$ in (\ref{q4}) and (\ref{q5}), respectively.
Eigenfrequencies, and eigenfunctions
were obtained by a shooting method described in Section
5.2 of Paper 1. Here we note that we integrated the standard full
set of four equations  describing adiabatic pulsations (eg.
Christensen-Dalsgaard 1998) to find these quantities for
relatively large values of  eigenfrequencies $\omega > 0.3-0.5$
depending on the particular model. These were then used to evaluate
the overlap integrals.
For smaller eigenfrequencies,  the Cowling
approximation is used to find the eigenfrequencies and the
expression (78) of Paper 1 was employed to find the overlap
integrals. This expression is equivalent to the original one presented
above provided the Cowling approximation is used to find eigenmodes.

We checked that
in the intermediate region for which  $\omega \sim 0.3-0.5$ both methods give
practically the same results.
 The advantage of  using the expression
(78) of Paper 1 in the low frequency limit is  the fact that the
integrand in this expression is less oscillatory in
comparison to the integrand in the original expression,  thus
allowing us to significantly increase accuracy of determination of
$\hat Q$ in this limit where the eigenfunctions contain  a
large number of nodes.

\begin{figure}
\begin{center}
\vspace{8cm}\includegraphics{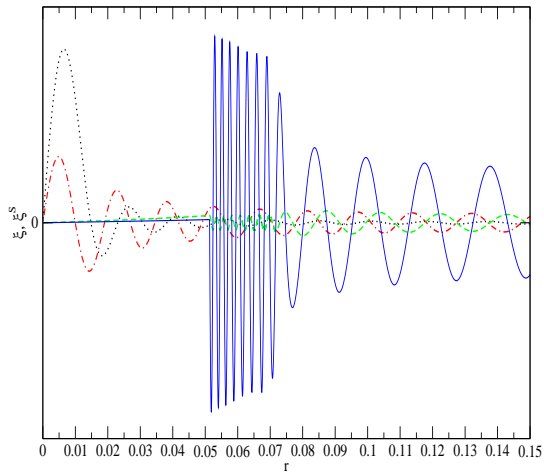}
\end{center}
\vspace{0.5cm} \caption{The radial dependence of the radial  and tangential components of the displacement vector,
$\xi$ and $\xi^{S},$ for typical eigenmodes
for models 2a and 2d. These are plotted in arbitrary units as functions of the radius $r,$ in the inner
region $0 < r < 0.15$.  Dotted and  dot dashed curves (black and red in the on-line version)
and  dashed and solid curves (green and blue in the on-line version) show $\xi$ and $\xi^{S}$,
respectively, for models 2a and 2d.
Fig. \ref{Fig2} indicates  that model 2a is mostly  radiative, while model 2d has a sharp
transition from the exterior  radiative region to the convective core which is  situated at $r_{c}\sim 0.05$. It is  seen
 that the eigenfunctions corresponding to the mostly  radiative model are smooth while those
 corresponding to the model with a convective core demonstrate a very sharp change of behaviour
in the vicinity of  $r\sim r_c$.}
\label{eigenfun}
\end{figure}

\begin{figure}
\begin{center}
\vspace{8cm}\includegraphics{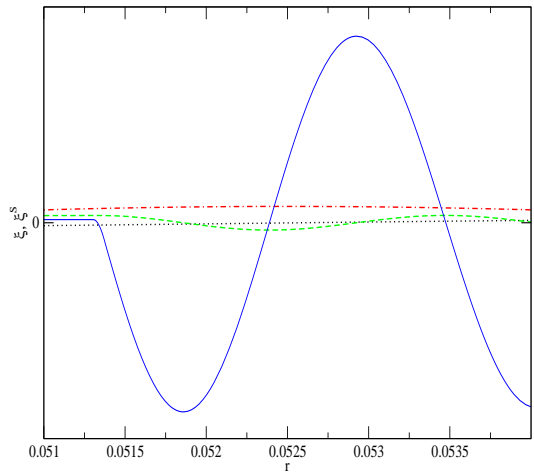}
\end{center}
\vspace{0.5cm} \caption{Same as Fig. \ref{eigenfun} but a region very close to
$r_c$ is shown. One can see that although the eigenfunctions corresponding to model 2d may
look discontinuous  in Fig. \ref{eigenfun}, in fact they are smooth. }
\label{eigenfun1}
\end{figure}

Models of massive stars have very rapid variations of
of $N^2$, see Figs \ref{Fig2}-\ref{Fig4}
in the transition  regions    between  radiative envelopes and  convective
cores,. Since the characteristic radial extent of
these variations can be of the order of the initial stellar model grid size
a  treatment of discontinuities may be needed.
 We prefer, however, to  avoid this situation in our approach.
To deal with this issue,  we worked with
computational grids which had  a much larger  number of grid
points than  were originally used to represent  the  structure of the stellar  models.
State variables were  interpolated
onto our more refined  grid as smooth functions.
There were then no discontinuities in the eigenfunctions
on the refined grid, see Figs \ref{eigenfun}, \ref{eigenfun1}.

\begin{figure}
\begin{center}
\vspace{8cm}\includegraphics{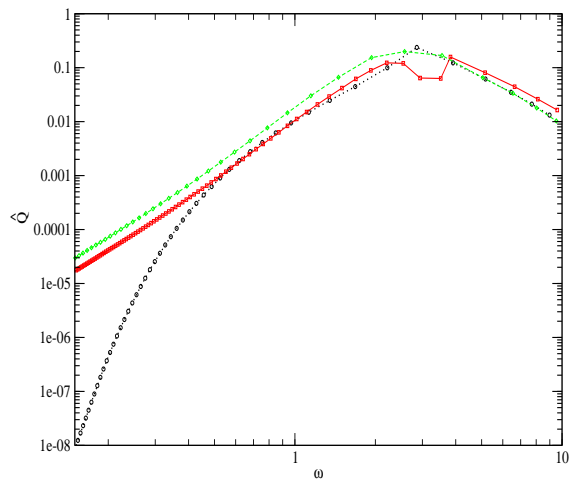}
\end{center}
\vspace{0.5cm} \caption{ The overlap integrals $\hat Q$ as
functions of mode eigenfrequency $\omega $ for models 1a and 1b
plotted with  dashed (green in the on-line version) and solid curves
(red in the on-line version), respectively. The black dotted curve
plots the overlap integrals for  a polytrope with $n=3.$
Note that for low frequencies,  these are much smaller than the ones corresponding to
stellar models with realistic structure. Symbols show the positions of
eigenfrequencies, with diamonds, squares and circles representing
the results for models 1a, 1b and the polytropic model, respectively.
The smooth curves are interpolated through these.} \label{Fig5}
\end{figure}
\begin{figure}
\begin{center}
\vspace{8cm}\includegraphics{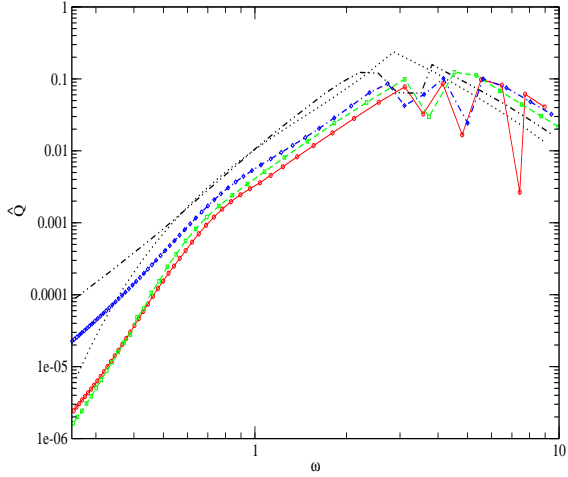}
\end{center}
\vspace{0.5cm} \caption{As for  Fig. \ref{Fig5} but for models
with $M_*=1.5M_{\odot}$.  Dot dashed (blue in the on-line version), dashed (green
in the on-line version)
and solid (red in the on-line version) curves are
for models 1.5a, 1.5b and 1.5c, respectively. We show the results for models
1p and 1b,  plotted as black dotted and dot  dot dashed curves respectively, for comparison.
Diamonds, squares and circles show positions of particular eigenfrequencies for models
 1.5a, 1.5b and 1.5c, respectively. }
\label{Fig6}
\end{figure}
\begin{figure}
\begin{center}
\vspace{8cm}\includegraphics{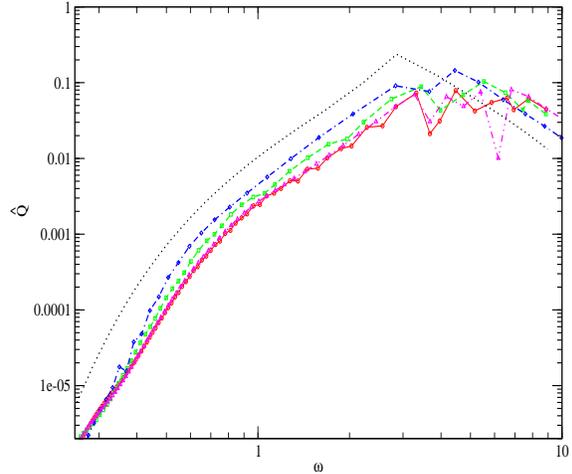}
\end{center}
\vspace{0.5cm} \caption{As for Fig. \ref{Fig5} but for models
with $M_*=2M_{\odot}$.  Dot dot dashed (magenta in the on-line version),
dot dashed (blue in the on-line version), dashed (green in the on-line version) and
solid (red in the on-line version) curves are for models 2a, 2b, 2c and 2d, respectively.
Triangles, diamonds, squares and circles show positions of particular eigenfrequencies
for models 2a, 2b, 2c and 2d, respectively.
The dotted curve gives the results for  model 1p. } \label{Fig7}
\end{figure}
\begin{figure}
\begin{center}
\vspace{8cm}\includegraphics{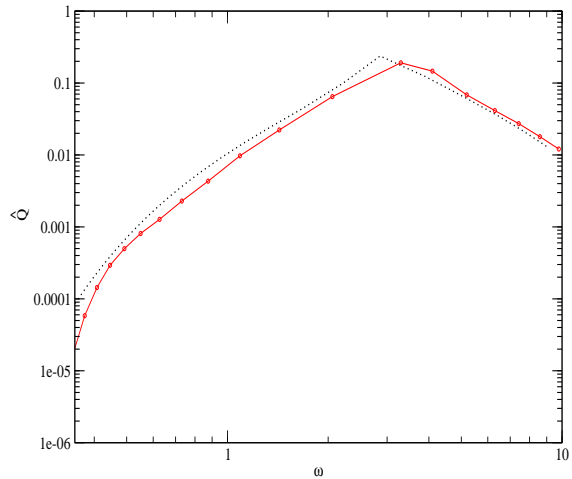}
\end{center}
\vspace{0.5cm} \caption{Same as Fig. \ref{Fig5} but for model 5a
of a young star with $M_*=5M_{\odot}.$   The solid (red in the on-line version )
curve and circles show the results calculated for this model. The  dotted curve is
for  model 1p.} \label{Fig8}
\end{figure}

The results of calculations of  overlap integrals are shown in Figs.
\ref{Fig5}-\ref{Fig8}.
 In Fig. \ref{Fig5} we show the results for
models 1p, 1a and 1b. Since the polytropic model 1p has been
discussed extensively elsewhere (eg. Press $\&$ Teukolsky 1977, Lee $\&$ Ostriker
1986, Ivanov $\&$ Papaloizou 2004 and references therein) and overlap
integrals corresponding to models 1a and 1b are discussed in
detail in Paper 1, here we only mention that the polytropic model
has much smaller overlap integrals when $\omega < 0.4$. This is
due to the fact that $\hat Q$ corresponding to Sun-like stars has
contributions arising from  the presence of a  convective envelope.
These decay  as a power of $\omega $ in the limit $\omega
\rightarrow 0,$  while the overlap integrals of the polytropic model
may be shown to decay faster than any power of
$\omega.$  Note that the overlap integrals for the solar model have also been
calculated recently by Weinberg et al (2012). Our calculations for the model 1b
agree quite well with their  results. It is  of interest to note
that the overlap integrals for  this model are not monotonic at large frequencies.
This effect is even more prominent for the models with $M_*=1.5M_{\odot}$ and $2M_{\odot}$ discussed
below, where several peaks in the values of $\hat Q$ at values of frequencies
 corresponding to pressure
modes are observed, see Figs. \ref{Fig6} and \ref{Fig7}. Since this effect is not
important for our purposes we do not discuss it here. However, we would like to mention
that it appears to be rather generic, e.g. it is present in the dependence of $\hat Q$ on
$\omega $ in a model of red giant star, see Fuller et al (2012).

Fig \ref{Fig6} shows  results for models 1.5a-1.5c.
For the  range of frequencies plotted,  only model 1.5a has  overlap
integrals larger than those of the polytrope at small values of
$\omega.$  As discussed above this is because of the fact that
this model has a rather extended convective envelope. From our
analytical theory developed in Paper 1,  it follows that in the  case of
model 1.5a,  the contribution determined by the presence of this
region should be roughly three times smaller than that arising
for   Sun-like stars. As seen from \ref{Fig6},  this
is confirmed by our numerical results.
More evolved models 1.5b
and 1.5c have rather small overlap integrals in the range of
frequencies plotted.  They are even smaller than  those  of the  polytropic
star. This indicates that tidal interactions determined by  the excitation of
eigenmodes in the shown range of frequencies are relatively weak
for these models\footnote{Note that
 the strength of tidal interactions
  also depends    on  the ratio of  the central and  stellar mean  density, see
below}.

Results for  stars of  mass $M_*=2M_{\odot}$ are  shown
in Fig. \ref{Fig7}. This case is rather similar to the previous
one.  However,  the more massive stars do not have well
pronounced convective envelopes and, therefore, their overlap
integrals are rather small for the range of frequencies shown,  being
several times smaller than those  for a  polytropic star. Finally,
in Fig. \ref{Fig8} the overlap integrals of a young star with
$M_*=5M_{\odot}$ are shown.  They are rather similar to,  though
slightly smaller than those  of a  polytropic star for the
range of frequencies shown. This means that in this frequency range the
contribution determined by the presence of a convective core
is probably  not  seen.

\subsection{Rotational splitting coefficients}
\begin{figure}
\begin{center}
\vspace{8cm}\includegraphics{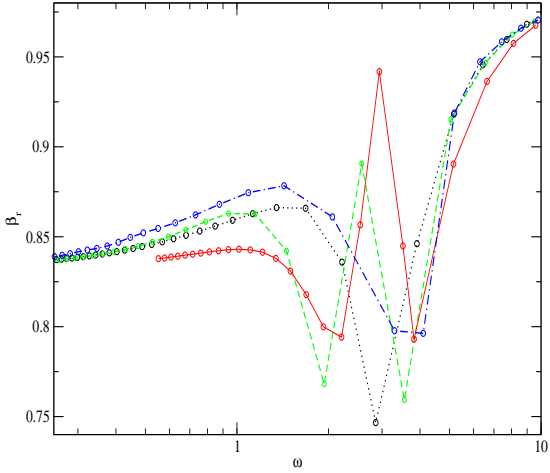}
\end{center}
\vspace{0.5cm} \caption{The rotational splitting coefficients
$\beta_r$ as functions of $\omega $ for models with
$M_*=M_{\odot}$ and $M_*=5M_{\odot}$. Dotted, solid, dashed and
dot dashed curves are for models 1p, 1b, 1a and 5a, respectively.
Open circles show positions of numerically calculated eigenfrequencies.} \label{b1}
\end{figure}
\begin{figure}
\begin{center}
\vspace{8cm}\includegraphics{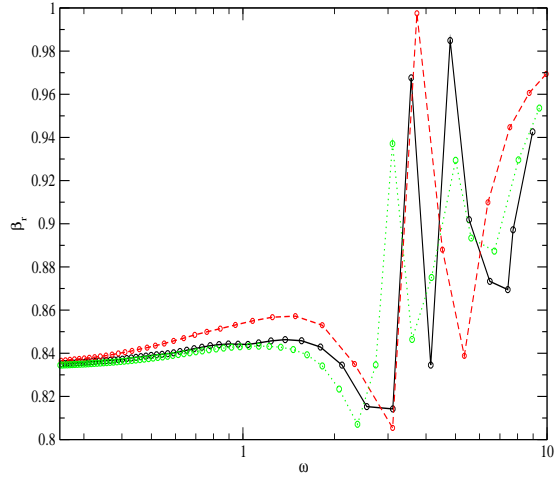}
\end{center}
\vspace{0.5cm} \caption{Same as Fig. \ref{b1} but for
$M_*=1.5M_{\odot}$. Solid, dashed  and dotted curves are for
models 1.5c, 1.5b and 1.5a, respectively.} \label{b1.5}
\end{figure}
\begin{figure}
\begin{center}
\vspace{8cm}\includegraphics{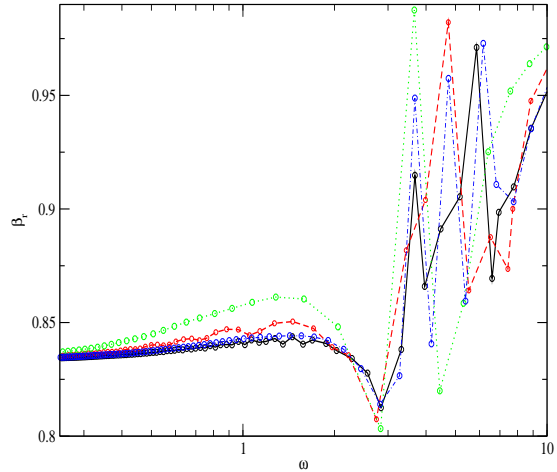}
\end{center}
\vspace{0.5cm} \caption{Same as Fig. \ref{b1} but for
$M_*=2M_{\odot}$. Solid, dashed, dotted and dot dashed curves are
for models 2d, 2c, 2b and 2a, respectively.} \label{b2}
\end{figure}

When the  rotation frequency of a star, $\Omega $, is
small in comparison to its natural frequency: $\Omega \ll
\Omega_*$ one may treat effects due  rotation in a simplified  manner
(eg. Lai 1997, Ivanov $\&$ Papaloizou 2011). From first order perturbation theory,
 the  eigenfrequencies,
$\omega_{j}$, are shifted with respect to their values
for  a non-rotating star, $\omega_{0,j}$, by an amount
proportional to $\Omega $. We take this shift into account but
assume that the overlap integrals are unchanged.

In the inertial frame we have (see eg.
Christensen-Dalsgaard 1998)
\begin{equation}
\omega_j=\omega_{0,j}+m\beta_{r}\Omega, \label{eq0}
\end{equation}
where $m=0,\pm 1, \pm 2 $ is the azimuthal number and $\beta_r$ are
dimensionless coefficients determining  the magnitude of the rotational splitting.
Note that when the rotation axis is
perpendicular to the orbital plane,  terms with  $m=\pm 1$
do not contribute to  the energy and angular
momentum transfer  as a result of a  periastron flyby.
The rotational
splitting coefficients $\beta_r$ can be expressed as a ratio
of two integrals involving the components of the mode  Lagrangian displacement.
When $\omega_{0,j} \gg 1$ they are
close to unity and when $\omega_{0,j} \ll 1$ they tend to $1-1/(l(l+1)
={5/ 6}$ for spherical harmonic index,  $l=2,$  (eg. Christensen-Dalsgaard 1998). In the
intermediate range of eigenfrequencies these integrals have to  be
evaluated numerically. We calculate them for our stellar models
and illustrate  them  for models 1p, 1a, 1b and 5a in Fig. \ref{b1}.  Results for
models 1.5a - 1.5c are illustrated in Fig. \ref{b1.5}, and
for models 2a-2d in  Fig. \ref{b2}. One can see from these figures  that the dependence
of $\beta_r$ on $\omega_{0,j}$ is not  necessarily monotonic,  as was also  found for
of the overlap integrals.

From the  results presented
below,  we find that use of  the perturbative
description of the influence of rotation on tides described above yields results that
agree quite well with those obtained from a  more
accurate approach based on the traditional approximation (Unno et al. 1989), even for quite large stellar
angular velocities $\Omega \sim 0.4.$ This is the case
when either the stellar rotation is retrograde with respect to
that of the orbital motion, or prograde with respect to the orbital motion,
but with the angular frequency being smaller in magnitude than
approximately the value  corresponding to pseudosynchronization,
for which there is zero angular momentum transfer.

\subsection{The overlap integrals in the traditional approximation}

\begin{figure}
\begin{center}
\vspace{8cm}\includegraphics{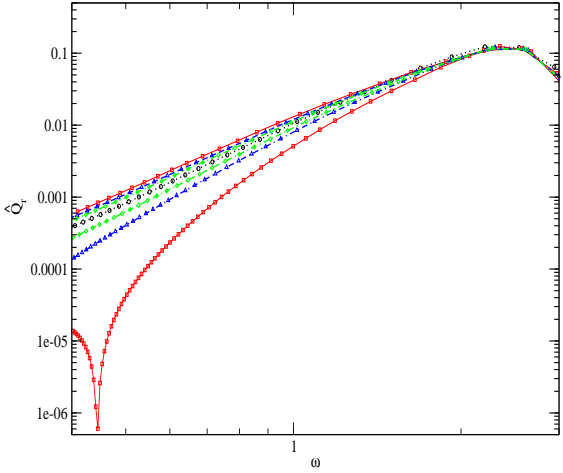}
\end{center}
\vspace{0.5cm} \caption{We show the radial contribution to
the overlap integrals $\hat Q_r$ for model 1b of the
present day Sun as functions of eigenfrequency  $\omega$ for different values of $\Omega.$
 The cases of $|\Omega|=0.42$, $0.21$, $0.11$ and the non-rotating case are shown
using  solid (red in the on-line version), dashed (blue in the on-line version),
dot dashed (green in the on-line version) and black dotted curves, respectively.
The curves of the same type with smaller (larger) values of $\hat Q_r$ for a given value of 
$\omega$ are calculated for retrograde (prograde) directions of rotation with respect to 
the pattern rotation associated with the forcing potential.
The curves corresponding to the three cases with prograde rotation almost coincide. 
Symbols show the positions of eigenfrequencies  with squares, triangles, diamonds
and circles corresponding to $|\Omega|=0.42$, $0.21$, $0.11$ and $0$, respectively.}\label{Qr1}
\end{figure}

\begin{figure}
\begin{center}
\vspace{8cm}\includegraphics{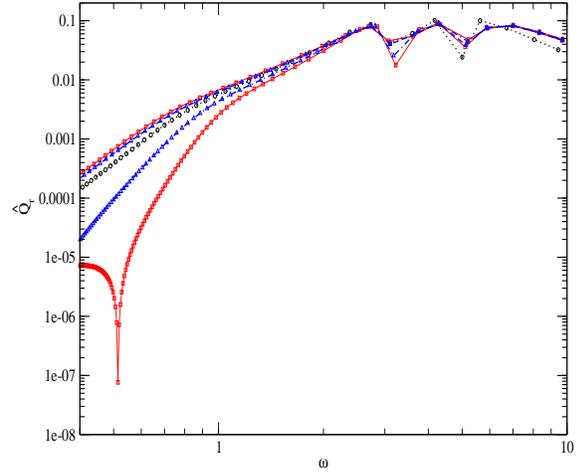}
\end{center}
\vspace{0.5cm} \caption{ As in  Fig. \ref{Qr1} but for model 1.5a.
The cases $|\Omega|=0$, $0.25$ and $0.5$  are plotted
using black dotted,  dot dashed (blue in the on-line version)  and  solid  (red in the on-line version)
curves, respectively. The curves corresponding to the three cases with prograde rotation almost 
coincide. Squares, triangles and
circles show positions of eigenfrequencies for  $|\Omega|=0.5$, $0.25$ and $0$,
respectively.}\label{Qr1.5}
\end{figure}

As explained in Paper 1 and above, when the traditional approximation is used
the overlap  integrals $\hat Q$ can be represented as products of 'angular' and 'radial'
contributions. Thus $\hat Q=\alpha \hat Q_{r}$, where the angular contribution $\alpha $ is
given in Fig. 1 of Paper 1. The radial contribution,
$\hat Q_r,$ is illustrated in Fig. \ref{Qr1} and
\ref{Qr1.5}, for models 1b and 1.5a, respectively.  Note that the curves corresponding
to  the largest  retrograde rotation  are not monotonic, with the overlap integrals having
a pronounced  minimum at some value of the  eigenfrequency. This effect is explained in Paper 1. Namely,
as follows from the discussion above,  the values of the radial contributions  to  overlap integrals for stars
with extended convective envelopes are mainly determined by contributions coming from
the convective  envelope and from the vicinity of the base of the  convective zone. It can be shown
(see Paper 1, equation (111)) that these contributions are proportional to the factor
$(1-30/\Lambda (\nu ))$, where we recall  that $\Lambda $ is the eigenvalue  associated with acceptable solutions
of the Laplace
tidal equation and $\nu=2\Omega/\omega $. It turns out that in case of retrograde rotation,
the value of $\Lambda $ corresponding to the frequencies, where this minimum is observed,
is approximately equal to thirty, and, therefore, the main contributions to the overlap
integrals are strongly suppressed.

We use below the traditional approximation described in detail in Paper 1
to calculate the energy and angular momentum transfer for model 1b as a result of a flyby
of a perturber on a parabolic orbit for
$\Omega =\pm 0.11$, $\pm 0.21$ and $\pm 0.42$ as well as for model 1.5a for
$\Omega =\pm 0.25$ and $\pm 0.5$.
These are compared with results obtained by numerically solving the flyby problem
directly as an initial value problem.

\section{Transfers of energy and angular momentum arising from  parabolic encounters and the tidal capture problem}

In this Section we discuss the well known  problem of calculating the  energy and
angular momentum transferred to  the normal modes of a star  as a result of  a parabolic encounter with a
perturber treated as a point mass. We consider realistic stellar models.
In general, we assume that the stellar
rotation axis is  perpendicular to the orbital plane.  For
a  discussion of the case of a general inclination between the stellar rotation axis
and the orbital angular momentum vector, see Ivanov $\&$ Papaloizou
(2011).

The magnitude of the   energy transferred, $\Delta E$, depends on
whether it is calculated  in the inertial or rotating
frame.
In general, we have
\begin{equation}
\Delta E_I = \Delta E +\Omega \Delta L, \label{eq1}
\end{equation}
where $\Delta E_I$ and $\Delta E$  are the energy transferred
as calculated  in the inertial and rotating frame, respectively, $\Omega $ is
the stellar angular velocity,   and $\Delta L$ is the  amount of
angular momentum transferred.

It is convenient to introduce natural units for the  energy and angular
momentum transferred.  Thus, we   express $\Delta E_I$ and $\Delta E$ in
units of $E_*=Gm^2_p/( (1+q)^2 R_*)$, where $G$ is the
gravitational constant, $m_p$ is the mass of a perturbing body, and
$q=m_p/M_*.$  We express  $\Delta L$ in units of $L_*={q^2
(1+q)^{-2}}M_*\sqrt{GM_*R_*}$. We remark that all quantities in  equations (54) of Paper 1
can also be
represented in natural units \footnote{The coefficients
$A_{m}$ should be expressed in units of $Gm_p/(
R^3_p \Omega_p)$, where $R_p$ is the periastron distance and a
typical periastron passage frequency
$\Omega_p=\sqrt{GM_*(1+q)/ R^{3}_p}$.
 The overlap integral should
be in the units of $\sqrt{M_*}R_*$ while the eigenfrequencies and
stellar angular frequency are  in the units of $\Omega_*$}. Once the
ratio $\Omega /\Omega_*$ is specified,  the energy and angular
momentum transferred   expressed in  natural units are
functions of only one parameter (see eg. Press $\&$ Teukolsky
1977, Ivanov $\&$ Papaloizou 2004, 2007)
\begin{equation}
\eta=\sqrt{{1\over 1+q}{\left({R_p\over R_*}\right)}^3}=3.05\sqrt{\bar
\rho}P_{orb} ,\label{eq2}
\end{equation}
where we recall  that $R_p$ is
the periastron distance. The quantity  $\bar \rho$ is the ratio of  the mean
stellar density  to the  solar value, $\bar \rho
={R_{\odot}^3M_* /(R_{*}^{3} M_{\odot})},$ and $P_{orb}$ is orbital
period of a circular orbit which has  the same value of the  orbital angular
momentum as the  parabolic orbit under
consideration, expressed in units of one day.

 Assuming that tidal
evolution approximately conserves angular momentum (see, eg.
Ivanov $\&$ Papaloizou (2011) for a discussion of this
approximation) $P_{orb}$ characterises the orbital period of the  binary
system after the process of tidal circularisation is complete.
Note that when $q \ll 1$ as for dynamic tides induced in a central
star in exoplanetary systems,  the condition $\eta =1$ corresponds
to a grazing encounter with periastron distance equal to the
stellar radius. Thus, for these systems only $P_{orb} >
P_{crit}=0.325/\sqrt{\bar \rho}$ are possible.

We remark  that a number of authors (eg. Press $\&$ Teukolsky 1977,
Lee $\&$ Ostriker 1986, Giersz 1986, McMillan, McDermott $\&$ Taam 1987)
 express results in terms of another dimensionless quantity, $T{_2}(\eta )$ \footnote{We remark  that we
consider the quadrupole component of the forcing potential.}, which is related to the dimensionless energy
transfer, $\Delta E$,   through
\begin{equation}
T_2(\eta)=\eta^4\Delta E.
\label{eq2aa}
\end{equation}
We use equation (\ref{eq2aa}) to compare our results with whose obtained by  previous authors.

When $\Delta E_I$ and $\Delta E$ are expressed in natural
units,  their dependence on $\eta$ may be used to compare
the strength of tidal interactions of stars with different masses and
radii.
However, for systems with given orbital parameters and
$m_p$ it also depends on the average density being larger for
stars with smaller $\bar \rho$, mainly through the dependence of
$\eta $ on this quantity. Therefore, in a similar way to our previous
studies (Ivanov $\&$ Papaloizou 2004, 2007, 2011) we introduce the
tidal circularisation time
\begin{equation}
T_{ev}=15\left({M_*\over M_{\odot}}\right)\left({R_*\over R_{\odot}}\right)\left({M_{J}\over m_p}\right){1\over \Delta
E_{I}}\sqrt {a_{in}} ,\label{eq3}
\end{equation}
where $M_J$ is the mass of  Jupiter ( see equation (104) of Ivanov \& Papaloizou 2007).  From here on, it is implied  that
energy and angular momentum transfers are expressed in  natural
units, and $T_{ev}$ is expressed in years. Under the assumption that the energy
transfers arising from  consecutive periastron passages can be simply
added,  $T_{ev}$  gives a characteristic time scale for the  tidal
evolution of the semimajor axis of a highly eccentric orbit with initial semi-major axis, $a_{in}$,
in units of $10$~AU. We   stress that this time scale applies only to the initial stages of circularisation
and will be characteristic of the whole process, only if it can proceed efficiently enough at small eccentricities
(see Ivanov \& Papaloizou 2011 for a discussion). We set $m_p=M_{J}$ and $a_{in}=1$ hereafter,  generalisation to other values of
$m_p$ simply  follows from the form of  equation (\ref{eq3}).

It is convenient to use equation (\ref{eq3}) when considering tidal interactions
in systems containing exoplanets. However,  dynamic tides may be also important
for other problems, such as eg. the  tidal capture of stars to form binary systems in stellar clusters
(eg. Fabian, Pringle $\&$ Rees 1975,  Press $\&$ Teukolsky 1977). To characterise the strength
of tidal interactions in such a setting it is convenient to introduce 'the capture radius'
$R_{cap}$, defined by the condition that when the periastron distance for a tidal encounter
between two initially unbound stars is equal to $R_{cap},$
the  initial relative  kinetic energy of these stars, when they are very far apart, is equal to
the amount of energy transferred due to tidal interactions,
 i.e.
 \begin{equation}\Delta E_{I}^{(1)}+\Delta E_{I}^{(2)}={1\over 2}{M_*^{(1)}M_*^{(2)}\over
M_*^{(1)}+M_{*}^{(2)}}v_{rel}^2,
\end{equation}
 where the upper index, $i,$  denotes  quantities associated with  star $i,$
and $v_{rel}$ is the initial  relative velocity
 of the stars with respect to  each other.
 Assuming that the binary consists of two identical stars,
which rotate  in the same sense with respect to their orbital motion, we get

\begin{equation}
\Delta E_{I}(\eta_{cap})=3.125\cdot 10^{-4}v_{*}^2, \quad v_{*}=\sqrt{({M_{\odot}\over M_{*}}{R_*\over R_{\odot}})}{v_{rel}\over 10km/s},
\label{eq4}
\end{equation}
and $\eta_{cap}$ is related to $R_{cap}$ through equation (\ref{eq2}) with $R_p=R_{cap}$.

\subsection{Energy and angular momentum transfer as a result of a parabolic encounter
 from direct numerical solution of the linear initial value problem}\label{direct}
We have  calculated the energy and angular momentum
transferred to a star as a result of an encounter with a perturber on
a parabolic orbit by solving the linear initial value problem directly. We refer to this procedure as the direct numerical
approach.
 The method  adopted follows from that described in
Papaloizou \& Ivanov (2010) and Ivanov \& Papaloizou (2011).
The equations solved are (36)-(41) of Ivanov and Papaloizou (2011)
with the following modifications. In that work a polytrope of
index $n=3$ and  a constant  adiabatic index $\gamma=5/3$ was considered.
However,  here we consider realistic stellar models for which   this varies.
Accordingly the quantity $P^{1/\gamma}/\rho\nabla (P'/P^{1/\gamma})$  in equation (36) of Ivanov and Papaloizou (2011)  was replaced
by $F_{ad}/\rho\nabla (P'/F_{ad}),$ where
\begin{equation}
F_{ad}= \int^P_{P_s} \frac{1}{\Gamma_1 P} dP,
\end{equation}
where $\Gamma_1 = (d\ln P/d\ln\rho )_{adiabatic}$ and  $P_s$ is the surface boundary
or photospheric pressure.
This quantity is readily obtainable for the models provided by numerical integration.
Elsewhere in equations (36)-(41) of Ivanov and Papaloizou (2011), $\gamma $ was replaced by $\Gamma_1.$

The presence of regions with negative $N^2 \equiv \omega^2_{BV}$
gave rise to linearly  convectively unstable eigenmodes which would ultimately dominate the solution.
In order to remove such modes, as long as the density gradient was negative,  we redefined $\Gamma_1$ in such regions such that $N^2\rightarrow 0$
there. This procedure amounts to stating that during linear perturbation,
the relationship between $P$ and $\rho$ in these regions is maintained.
 Here we are  adopting  the common approximation that the layers are effectively  adiabatically stratified (eg. Ogilvie \& Lin 2007).
This is equivalent to the condition that the convective, or frictional time scale (cf. Zahn 1977) be significantly longer than the rotation period,
which is expected to be satisfied for the cases we consider.

In some models, there were small low density regions where the density gradient became positive
 ( see discussion in Section \ref{Modeldesc} above).
For these regions, again $N^2$ was set to zero, but $\Gamma_1$ was not allowed to become negative,
instead being set to be the largest positive value attained on the grid from the first procedure.
In this way an incompressibility condition is approached.
However, as the values of the density and pressure were very small, this did not lead to  numerical
difficulties or, as shown by numerical tests,  affect results significantly. As in our previous work, most simulations were carried
out on a $200\times 200$ numerical grid with $m=2$ which gives the dominant contribution.
Resolution tests were carried out by doubling the resolution to  $400\times 400$
and as in our previous work showed good convergence in these cases  with variable $\Gamma_1.$


\subsection{Non-rotating stars}

\begin{figure}
\begin{center}
\vspace{8cm}\includegraphics{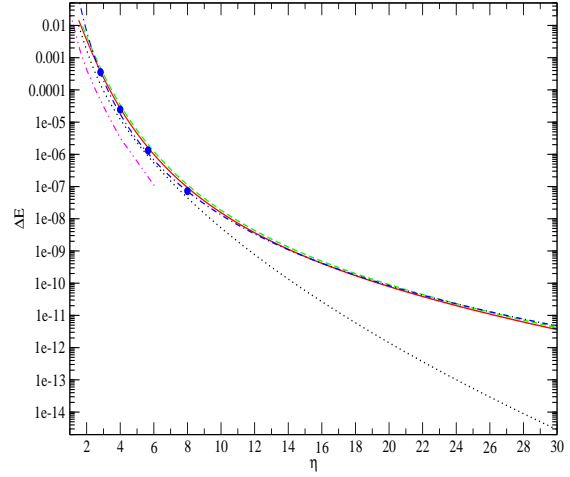}
\end{center}
\vspace{0.5cm} \caption{Energy transferred to the normal modes of a
non-rotating star with $M_*=M_{\odot}$ as the result of  a parabolic flyby of
a perturber of mass $m_p=M_J$, expressed in units of $E_*$, as a
function of the parameter $\eta $. The solid, dashed, and dotted
curves are for models 1b, 1a and 1p, respectively. The
dot dashed curve shows results  calculated using the  purely analytic expressions
for the mode eigenfrequencies and overlap integrals obtained in Paper 1 for model
1b. This almost coincides with the solid line. The  dot dot dashed curve  shows the 
result of Giersz (1986). Circles show
the energy transfer calculated for model 1b using the direct
numerical approach.} \label{Fig9}
\end{figure}

\begin{figure}
\begin{center}
\vspace{8cm}\includegraphics{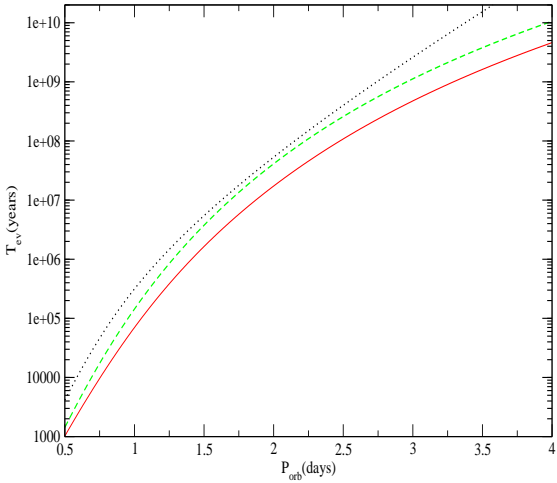}
\end{center}
\vspace{0.5cm} \caption{The evolution time $T_{ev}$ as a function
of the orbital period after circularisation, $P_{orb}$, for the
same models as illustrated in Fig. \ref{Fig9}. Again, the solid, dashed, and
dotted curves are for models 1b, 1a and 1p, respectively.}
\label{Fig10}
\end{figure}

\begin{figure}
\begin{center}
\vspace{8cm}\includegraphics{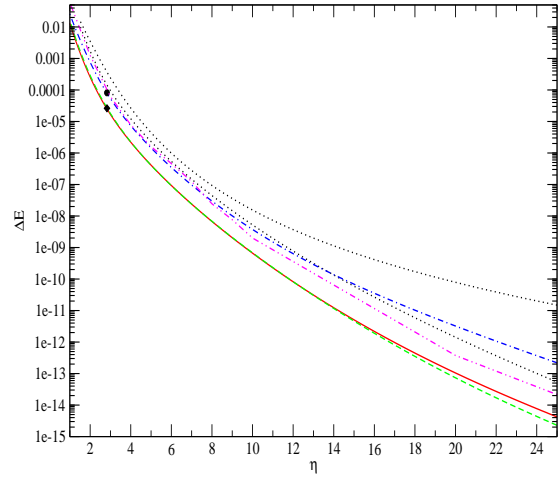}
\end{center}
\vspace{0.5cm} \caption{Same as in Fig. \ref{Fig9} but for the
models with $M_*=1.5M_{\odot}$. The two dotted curves  are for our
'reference' models 1p and 1b.   The polytropic star, model 1p, has a
smaller value of $\Delta E$ for  a given value of $\eta.$ Solid,
dashed and dot dashed curves are for models 1.5c, 1.5b and 1.5a,
respectively. As in  Fig. \ref{Fig9},
symbols indicate the amount of energy transferred  that was obtained adopting  the
direct numerical approach.  The   circle and square
are  for models 1.5a and 1.5c, respectively. The dot dot  dashed curve shows  the results  of
McMillan, McDermott $\&$ Taam 1987 for a population II model.} \label{Fig11}
\end{figure}

\begin{figure}
\begin{center}
\vspace{8cm}\includegraphics{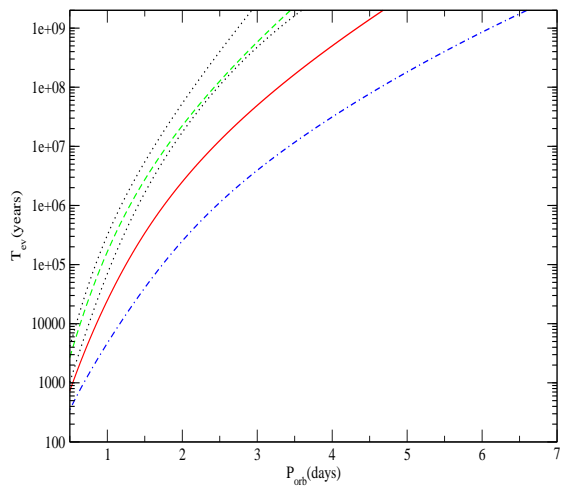}
\end{center}
\vspace{0.5cm} \caption{The evolution time $T_{ev}$ for models
with $M_*=1.5M_{\odot}$ as a function of $P_{orb}$. Curves of
a given  type plotted in Fig. \ref{Fig11} and Fig. \ref{Fig12} correspond to
the same models. } \label{Fig12}
\end{figure}

\begin{figure}
\begin{center}
\vspace{8cm}\includegraphics{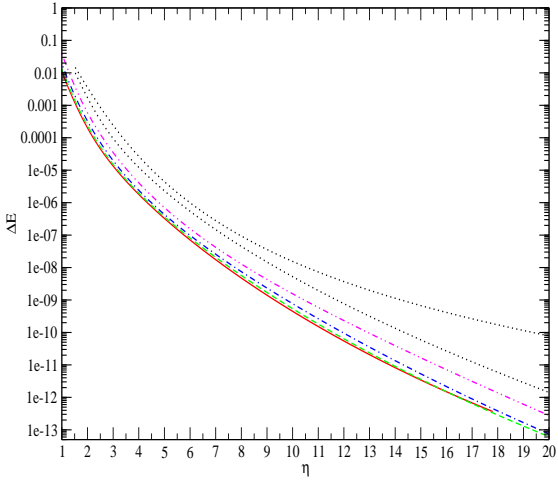}
\end{center}
\vspace{0.5cm} \caption{Same as in Fig. \ref{Fig11} but for the
models with $M_*=2M_{\odot}$. Solid, dashed, dot dashed and dot
dot dashed curves are for models 2d, 2c,  2b  and 2c respectively,
the dotted curves are the same as in Fig. \ref{Fig11}.}
\label{Fig13}
\end{figure}

\begin{figure}
\begin{center}
\vspace{8cm}\includegraphics{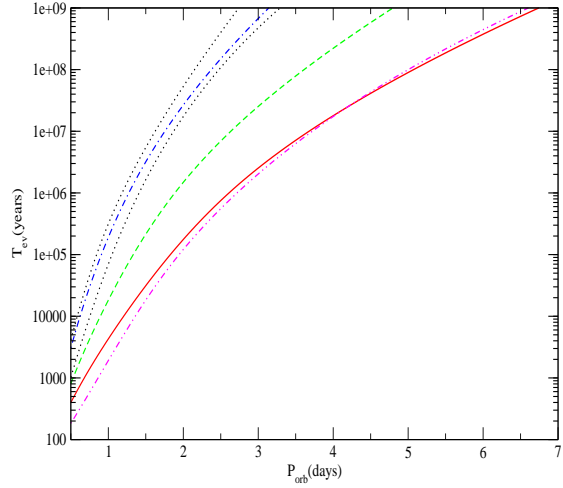}
\end{center}
\vspace{0.5cm} \caption{The evolution time $T_{ev}$ for models
with $M_*=2M_{\odot}$ as a function of $P_{orb}$. Curves of the
same type in Fig. \ref{Fig13} and Fig. \ref{Fig14} correspond to
the same models. } \label{Fig14}
\end{figure}

\begin{figure}
\begin{center}
\vspace{8cm}\includegraphics{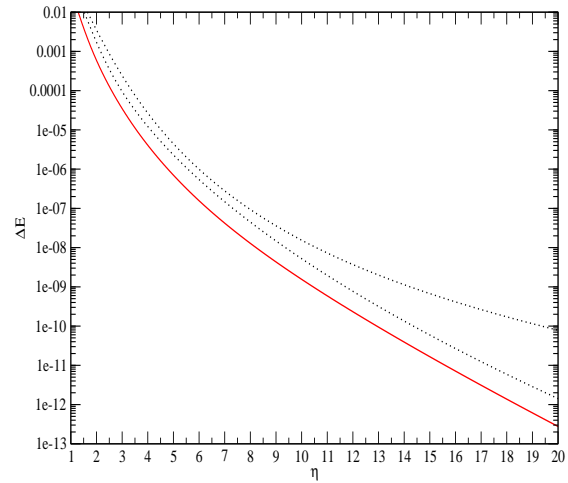}
\end{center}
\vspace{0.5cm} \caption{The energy transfer $\Delta E$ for the
model 5a with $M=5M_{\odot}$ shown by the solid curve, the dotted
curves show our reference models 1p and 1b as in the previous
Figs.} \label{Fig15}
\end{figure}

\begin{figure}
\begin{center}
\vspace{8cm}\includegraphics{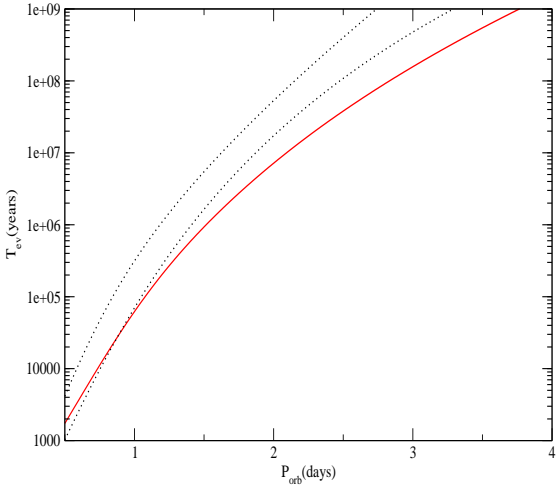}
\end{center}
\vspace{0.5cm} \caption{Same as Fig. \ref{Fig15} but for the
evolution time $T_{ev}$.} \label{Fig16}
\end{figure}

In this Section we consider non rotating stars, accordingly $\Omega=0.$ In this case
$\Delta E_{I}=\Delta E$. The results of
numerical calculations of $\Delta E$ and $T_{ev}$ for the  stellar
models presented in table \ref{t1} are shown in Figs.
\ref{Fig9}-\ref{Fig16}.

In Fig \ref{Fig9}
we show the dependence of $ \Delta E$ on $ \ eta $
calculated for our models of Sun-like stars, 1a and 1b, together
with the same quantities  calculated for a polytropic star with solar
mass and radius.  Circles show results
obtained from the direct numerical approach. One can see
 that the approach based on  normal mode calculation gives
practically the same energy transfer as the direct numerical approach. A small
deviation for  $\eta=8$  are probably due to the effect of numerical
diffusion  and finite integration time. As seen from  Fig.\ref{Fig9}, the
energy transfers for  realistic models are significantly larger
than those for  the polytropic model for large  enough values
of $\eta .$  Clearly, this is due to much larger values of the
overlap integrals for  the realistic models at small
eigenfrequencies, see Fig. \ref{Fig5}.
 Interestingly, the energy
transfers for models 1a and 1b are quite close to each other
regardless of the fact that the overlap integrals corresponding to
model 1a are larger than those of model 1b. This is explained
by observation that the number of eigenmodes contributing  to the
tidal interaction is larger in the case of model 1b. This
compensates for smaller values of the overlap integrals.

The dot dashed curve shown in  Fig. \ref{Fig9}  is obtained using  the formalism
described in  paper 1 for
model 1b. This curve is calculated by purely analytic means. We see that
there is quite good agreement between the analytic and numerical results. In the
range $2< \eta < 10$ the deviation is at most about  40 per cent, when $\eta > 10,$
corresponding curves practically coincide. The dot dot dashed curve in the same figure
shows the result of Giersz (1986) for a solar model. The energy transfer found  Giersz (1986)
is significantly smaller than we  obtain in this paper. Since our results are obtained
using  three independent methods,  we believe  that the  Giersz (1986) result underestimates  the energy
transfer, though the origin of the discrepancy is unclear. One possible explanation is  that
the  number of eigenmodes used in his calculation was too small.

From Fig. \ref{Fig10} it follows that the evolution time scale
$T_{ev}$ is smaller for  model 1b, than for model 1a,
for a given value of $P_{orb}$\footnote{  Note that
 our calculations for model 1a are not
realistic when $T_{ev} $ exceeds its age $\approx 1.7\cdot
10^{8}$ years. More realistic calculations of $T_{ev}$ should
employ a set of overlap integrals and eigenfrequencies calculated
for a grid of stellar models of  different ages.}.
The fact that
the model of present day Sun, model 1b, can be tidally excited  more
efficiently than the  corresponding  polytropic model, leads to the conclusion
that taking into account  realistic stellar models can
significantly increase estimates of the contribution of tides exerted on  the star for the
orbital evolution of Jupiter mass  exoplanets on highly eccentric
orbits.  The enhanced   tidal interaction can produce a significant change of the orbital semimajor
axis in a time  of less than $4\cdot 10^9$ years when $P_{orb}< 4.$
Note that this contribution can be further amplified for  stars
rotating in the opposite sense to that of  the orbital motion, as
discussed in Lai (1997), Ivanov $\&$ Papaloizou (2011) and below.
Furthermore, tides exerted on the star  become even more efficient for more
massive planets, see eg. Ivanov $\&$ Papaloizou (2004), (2007).
We also remark that a consequence of the above is that a significant component of the energy
liberated by orbital circularisation may be dissipated in the star rather than the planet,
thus  alleviating  the possibility of the  potential destruction of the planet
(see eg. the discussion in Ivanov \& Papaloizou 2004).

In Figs. \ref{Fig11} and \ref{Fig12} we show the energy transfer
$\Delta E$ and the evolution time $T_{ev}$ calculated for more
massive stellar models with $M_*=1.5M_{\odot}$. The dotted curves
on both Figs.  are for  our reference models, namely the  polytropic model 1p,
and the solar model 1b. As seen from Fig. \ref{Fig11} the energy
transfer expressed in the natural units is significantly smaller
for all the more massive models as compared to the
solar model. This is obviously because of the  smaller values of the
overlap integrals, see Fig. \ref{Fig6}. However, at large values of
 $\eta,$ model 1.5a gives  larger values for the  energy transfer than
the polytropic model. As discussed above, this is due to the
presence of rather extended convective envelope in  model 1.5a,
which results in larger values of the overlap integrals at small
eigenfrequencies as  compared  to models 1p, 1.5b, and 1.5c.

The symbols in Fig. \ref{Fig11} indicate
 the energy transfer obtained from  the direct numerical
approach  for models 1.5a and 1.5c. As seen from  Fig.\ref{Fig11},  the agreement between
this method and the normal mode  approach is excellent. The  dot dashed curve  shows the results of
 McMillan, McDermott $\&$ Taam (1987) for a Population II $1.5M_{\odot}$ star.
At  sufficiently large values of $\eta,$ the energy transfer for  their model is significantly
smaller than that for  model 1.5a,  but larger than that for  models 1.5b and 1.5c.
Unfortunately,  McMillan, McDermott $\&$ Taam (1987) did not provide details of their model.
Thus it is unclear as to  whether this behaviour  is a consequence of
the form  of the Brunt - V$\ddot {\rm a}$is$\ddot {\rm
a}$l$\ddot {\rm a}$ frequency, as  is indicated from consideration of our models.

However, it is important to point out that, for a given $P_{orb},$  the circularisation times, $T_{ev}$, corresponding to the  more
massive models are significantly smaller than those for  the
Sun-like models. This is due to the fact that the mean  density
of the more massive models is significantly smaller than the mean
density of the Sun-like models, see table \ref{t1}, which leads to
smaller values of $\eta $ for the more massive  models for  a given value
of $P_{orb}$, see equation (\ref{eq2}). In particular,  model 1.5c
having the largest age $\sim 1.6\cdot 10^{9}$ years,  has
$T_{ev}$ of the order of or smaller than its age, for planets
having $P_{orb} < 4$. The increase of the efficiency of the tidal interaction
for model 1.5c,  as compared to models 1.5b and 1b,  can also be viewed as a consequence of the expansion of the
star that takes place as a result of evolution.

Models with $M=2M_{\odot}$ and $M=5M_{\odot}$ are qualitatively
similar to models 1.5b and 1.5c. In all cases the energy transfer
is even smaller that   for the polytropic model,
while the evolution times $T_{ev}$ are smaller than  those for
the reference models 1p and 1b, due to the smaller mean  densities
of the massive stars. Note that the energy transfers calculated
for models 2c and 2d are very close to each other. It is also
interesting to note that the value of the mean  density evolves
non-monotonically with time. For $M_*=2M_{\odot},$
model 2d with the greatest age $\sim 10^9$ years, this  is
smallest. This leads to possibility of significant evolution
of the semimajor axis,  with $T_{ev} < 10^9$ years, for exoplanets
orbiting stars with  $M_*=2M_{\odot}$ and final orbital periods,
assuming the circularisation process can be completed of  $P_{orb} < 7$~days,
solely due to tides exerted on the star.

\subsection{Rotating stars in the traditional approximation}

In this Section we present results
 for  rotating models  1b and 1.5a within the framework of the traditional approximation
described in detail in Paper 1. We compare these  results to those
obtained from  the perturbative approach, where it
is assumed that the overlap integrals are not changed by rotation, and the mode eigenfrequencies  as viewed in the inertial frame
are shifted by a factor $m\beta_{r}\Omega $, as given by equation (\ref{eq0}). The perturbative
approach is described in more  detail in Ivanov $\&$ Papaloizou (2011) and  references therein.
Results are also compared to those obtained following the numerical approach,
that is by solving the encounter problem directly as an initial value problem
(see section \ref{direct}).
 In all of this it is assumed
that the rotation axis is perpendicular to the orbital plane, with both prograde and retrograde encounters with
respect to the direction of the stellar rotation
being  considered \footnote{For discussion of
the general case of an arbitrary inclination of the stellar rotational axis, see Ivanov $\&$ Papaloizou (2011).}.

\subsection{A comparison of eigenspectra obtained from  normal mode calculations  with those obtained from
the direct numerical approach}

\begin{figure}
\begin{center}
\vspace{8cm}\includegraphics{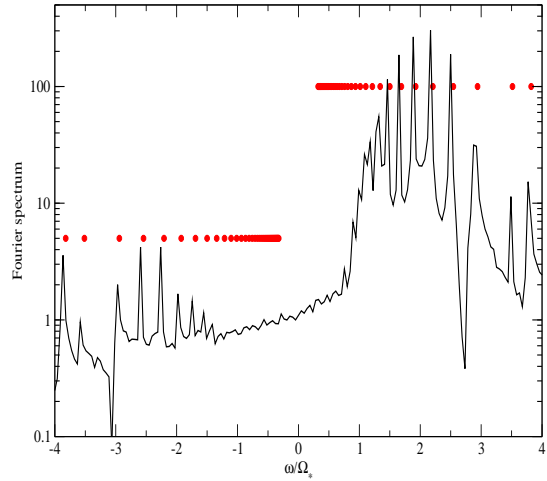}
\end{center}
\vspace{0.5cm} \caption{The amplitude of the time Fourier transform
 of  the  velocity component in the $\theta$ direction
evaluated at a characteristic  interior point  as a function of frequency $\omega.$
The case of non-rotating model 1b and $\eta=4\sqrt{2}$ is presented. Circles show the positions of eigenfrequencies calculated by
finding  the normal modes, assuming the traditional approximation directly (the normal mode method).}\label{S1}
\end{figure}

\begin{figure}
\begin{center}
\vspace{8cm}\includegraphics{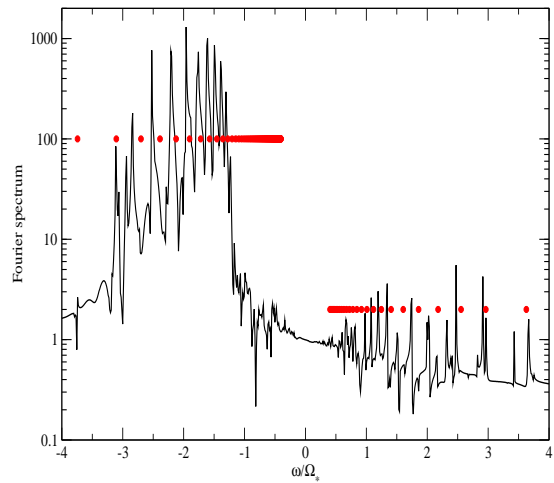}
\end{center}
\vspace{0.5cm} \caption{Same as Fig. \ref{S1} but for a retrograde tidal encounter with $\Omega=0.58$.}\label{S2}
\end{figure}

In order to check whether our direct numerical method and the normal modes approach agree with each other we consider
$\theta$-component of the perturbed velocity at a characteristic  position in the star as a function of time and make a Fourier transform
of the signal. The results are shown in Figs. \ref{S1} and \ref{S2} for a tidal encounter of a non-rotating star and
a retrograde encounter with $\Omega=0.58$, respectively. In the latter case positive and negative
values of the frequency $\omega$ correspond to perturbations propagating in the direction of stellar rotation and opposite to
this direction, respectively. The value of the amplitude scaling of the Fourier transform is arbitrary. Peaks in these Figs.
indicate the  approximate positions of free normal mode pulsations. Symbols show the positions of eigenfrequencies calculated
using  the normal mode approach.

As seen from Fig. \ref{S1},  in the case of the non-rotating star,  mainly positive eigenfrequencies are excited in the course
of tidal encounter. Positions of peaks are in  rather good agreement with the normal mode calculations for modes having
$|\omega| > 1$. Smaller values of $\omega $ are not well resolved on account of the finite time  duration of the run.

Contrary to the non-rotating case,  the retrograde tidal encounter mainly excites
eigenmodes with negative  eigenfrequencies, see Fig. \ref{S2}. Again, the  most prominent peaks approximately
in the range $-3 < \omega < -1$ are in a rather good agreement with the normal mode method. The eigenfrequencies found
from the normal mode method  are,
however, shifted towards $\omega=0$ with respect to the  positions of the corresponding peaks,
 the shift being  larger for eigenmodes with
larger absolute values of $\omega $. The situation is similar for modes with positive values of $\omega$
in the range $1 < \omega < 3.5$, but in this case
mode eigenvalues  are shifted towards larger values of $\omega$ with respect to corresponding  peak positions.
This disagreement
is possibly determined  by the fact that we use the Cowling and
traditional approximations to calculate the eigenspectrum in the normal mode approach.
Since smaller absolute  values of the
eigenfrequencies result in larger values of the associated transfers of energy and
angular momentum, we expect that the direct numerical approach
gives smaller values of these quantities in the case of retrograde encounters and larger values for
prograde encounters. This is indeed obtained in our calculations, see the discussion below.
Note too the absence of significant peaks in the inertial range for which $-1.16 < \omega < 1.16.$
This is where inertial modes associated with the convective envelope would be expected to show up.
However,  when confined to a spherical annulus inertial waves  may focus on to  wave attractors becoming singular
in the inviscid limit .  Then  corresponding discrete inviscid modes may not exist (eg. Ogilvie \& Lin  2004,  Ogilvie 2013),  but instead a continuous spectrum,
with the consequence that very  prominent resonant spikes are not seen in the response.
Nonetheless the  relatively small response in the inertial range indicates that inertial waves are not important in this
particular case.

\subsection{Energy and angular momentum transfer for rotating stars}

\begin{figure}
\begin{center}
\vspace{8cm}\includegraphics{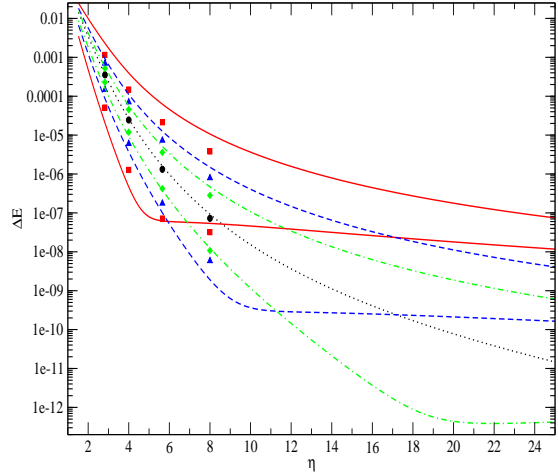}
\end{center}
\vspace{0.5cm} \caption{The energy transferred in the rotating frame expressed in natural units
as a function of  $\eta.$  Curves of different style correspond to different absolute values of the angular velocity
$\Omega$ and symbols indicate  results obtained using the direct numerical approach. See
the text for the allocation  of the  curves with different styles.} \label{Fig17}
\end{figure}

\begin{figure}
\begin{center}
\vspace{8cm}\includegraphics{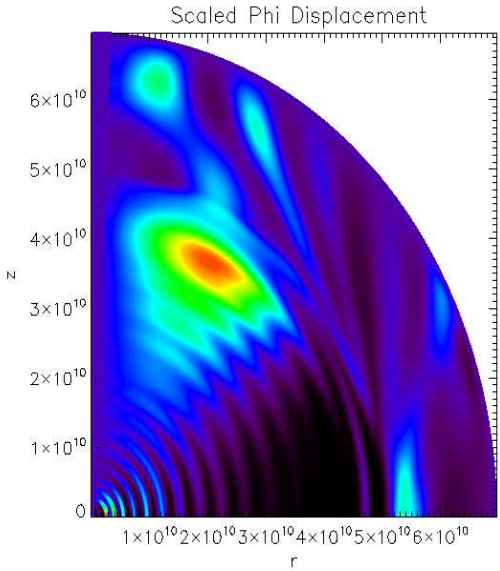}
\end{center}
\vspace{0.5cm} \caption{Illustration of inertial waves in the convective envelope for the
prograde  encounter  of model 1b with $\eta=8$ and  $\Omega=0.21.$
Contours of the product of the Lagrangian displacement in
the $\phi$ direction and $\sqrt{\rho}$ are shown in an upper quadrant at a time after the encounter is over and the energy transfer has
been completed. The presence of rotationally modified $g$ modes of order up to 15 can be seen in the radiative core while the convective envelope shows the presence of inertial waves
that show reflections as well as  graze the boundary with the radiative core.
The radial coordinate is expressed in $cm.$
  \label{Figsection}}
\end{figure}

\begin{figure}
\begin{center}
\vspace{8cm}\includegraphics{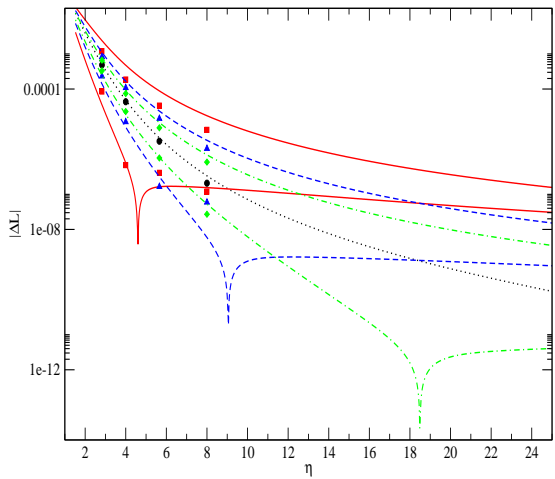}
\end{center}
\vspace{0.5cm} \caption{Same as Fig. \ref{Fig17}, but for the
absolute value of transferred angular momentum, $|\Delta L|$.
The locations where $\Delta L$ decreases dramatically in magnitude
correspond to pseudosynchronization (see text)} \label{Fig18}
\end{figure}

\begin{figure}
\begin{center}
\vspace{8cm}\includegraphics{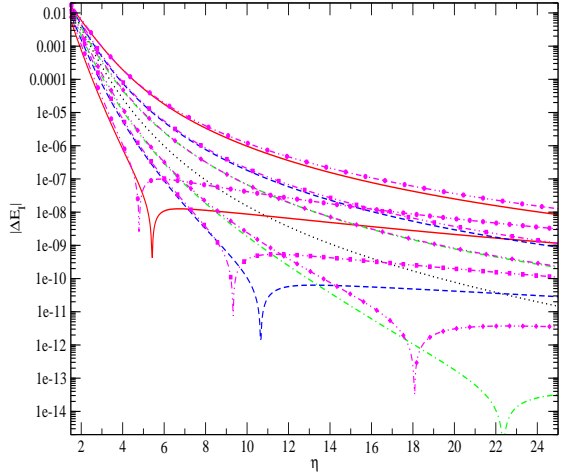}
\end{center}
\vspace{0.5cm} \caption{Same as Figs \ref{Fig17} and \ref{Fig18}, but for the
absolute value of energy in the inertial frame, $|\Delta E_{I}|$. We recall  that $\Delta E_{I}$ is related
to $\Delta E$ and $\Delta L$ through equation (\ref{eq1}). Additionally, we show the energy transfer calculated with the
framework of the perturbative approach by dot dot dashed lines. Different symbols on these lines correspond to different
rotation rates. } \label{Fig19}
\end{figure}

\begin{figure}
\begin{center}
\vspace{8cm}\includegraphics{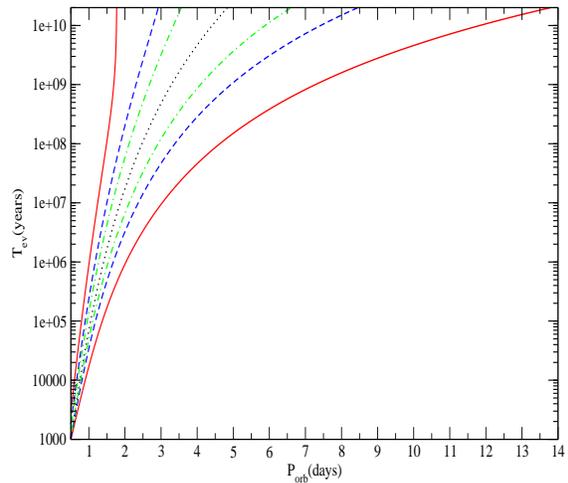}
\end{center}
\vspace{0.5cm} \caption{The evolution time $T_{ev}$  defined through  equation \ref{eq3} as a  function of the orbital period
after circularisation $P_{orb}$ for the rotating solar models. Curves of a given  style apply to
the same rotation rates as  in Fig. \ref{Fig17}.} \label{Fig20}
\end{figure}

Figs. \ref{Fig17}-\ref{Fig20}  are for model 1b. In these Figs. we show results obtained from the full numerical approach
applied to Sun-like stars having $|\Omega|=0.42$, $0.21$, $0.11$ and $0.$ These are
represented by solid, dashed, dot dashed and dotted
curves, respectively. In Figs. \ref{Fig17}-\ref{Fig19} the retrograde encounters have a larger value of the transferred
quantity   at a given
value of $\eta $ then the prograde encounters. The
squares, triangles, diamonds and circles in Figs.  \ref{Fig17}-\ref{Fig18} show
the corresponding results obtained using  the direct numerical approach, respectively for
$|\Omega|=0.42$, $0.21$, $0.11$ and $0$.

\begin{figure}
\begin{center}
\vspace{8cm}\includegraphics{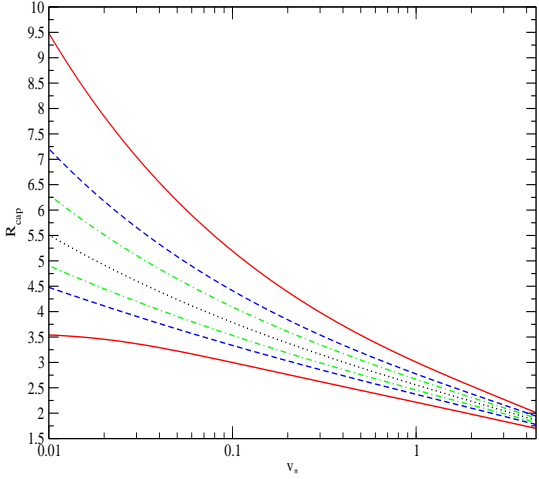}
\end{center}
\vspace{0.5cm} \caption{The capture radius $R_{cap}$ calculated according to equation (\ref{eq4}) for model 1b as a function of
a 'typical' relative velocity $v_{*}$. Curves of a given  style
apply to  the same rotation rates  as in  Fig. \ref{Fig17},
 with larger values of $R_{cap}$ for a given value of $v_{*}$ corresponding to
retrograde encounters.} \label{Fig21}
\end{figure}

\begin{figure}
\begin{center}
\vspace{8cm}\includegraphics{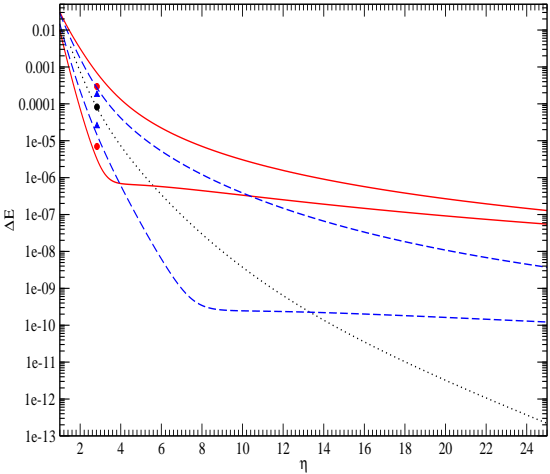}
\end{center}
\vspace{0.5cm} \caption{The energy exchange $\Delta E$ as a function of $\eta$ for a  rotating model 1.5a.
Solid curves and squares, dashed curves and triangles,  and finally  dotted curves and circles
correspond to $|\Omega|=0.5$, $0.25$ and $0$, respectively.} \label{Fig22}
\end{figure}

\begin{figure}
\begin{center}
\vspace{8cm}\includegraphics{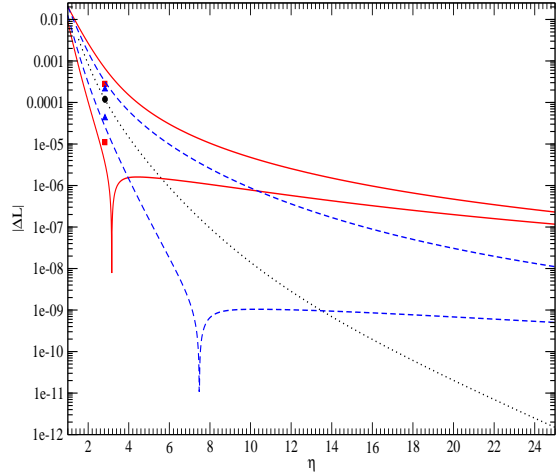}
\end{center}
\vspace{0.5cm} \caption{Same as Fig. \ref{Fig22} but the transfer of angular momentum $\Delta L$ is shown.} \label{Fig23}
\end{figure}

\begin{figure}
\begin{center}
\vspace{8cm}\includegraphics{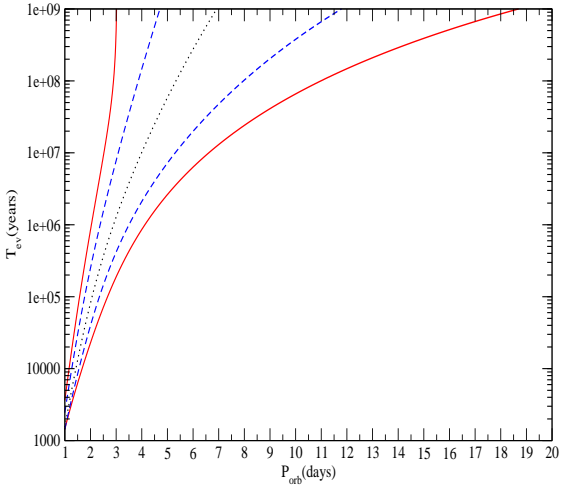}
\end{center}
\vspace{0.5cm} \caption{The evolution time scale $T_{ev}$ as a function of the orbital period $P_{orb}$ for model 1.5a.
Curves of a given   style represent the  same rotation rates as in Figs. \ref{Fig22} and \ref{Fig23}.
For a a given absolute value of $\Omega,$  the retrograde case corresponds
to a smaller value of $T_{ev}$ for a given $P_{orb}.$} \label{Fig24}
\end{figure}

\begin{figure}
\begin{center}
\vspace{8cm}\includegraphics{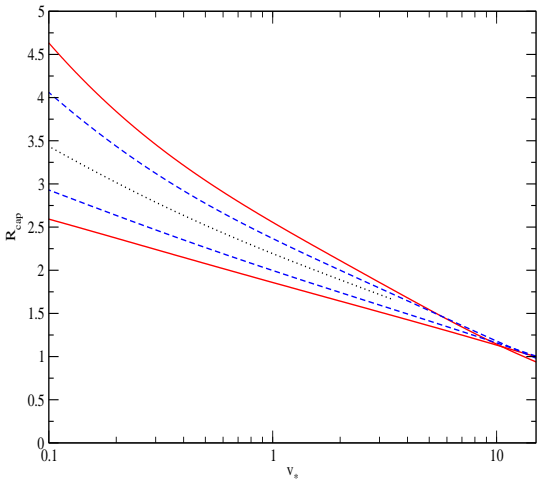}
\end{center}
\vspace{0.5cm} \caption{The radius $R_{cap}$ as a function of velocity $v_{*}$ for rotating 1.5a model. Curves of a given  style
apply as for Fig. \ref{Fig22}.
For a a given absolute value of $\Omega,$  the retrograde case corresponds
to a larger value of $R_{cap}$ for a given $P_{orb}.$
 \label{Fig25}}
\end{figure}

In Fig. \ref{Fig17} we show the transfer of energy in the inertial frame, $\Delta E$, versus $\eta$.  We see that
 retrograde encounters always have a larger value of $\Delta E$ than prograde encounters, for a given $\eta$,
with this effect being more extreme for larger absolute values of the angular velocity. The physical reason for  this behaviour
is discussed in detail in Lai (1997) and Ivanov $\&$ Papaloizou (2011). It is interesting to note that
 prograde encounters  with non zero angular velocity
produce larger values of $\Delta E$ than the case with $\Omega=0,$ when $\eta $ is large.
Within the framework of rotationally modified gravity modes under  the traditional approximation,
this is explained by noting  that in the limit of very large values of $\eta,$ the eigenmodes
mainly determining the value of $\Delta E$ propagate in retrograde direction with respect to the stellar
rotation in the rotating frame, but  at the same time, they propagate in the prograde direction in the
inertial frame, since the pattern speed  of these modes is smaller than the angular velocity of rotation.
 From the results of eg. Ivanov $\&$ Papaloizou 2011 it follows that
the energy transfer due to these modes is proportional to $ \eta^{-2}[\sum_i Q_i^2I^{2}_{2,-2}(y)]$,
where $y=\eta (2\Omega-|\omega_i|) $, the quantities $I_{l,m}(y)$ are discussed in Press $\&$ Teukolsky (1977),
see also Ivanov $\&$ Papaloizou (2007). The function $I^{2}_{2,-2}(y)$ has a maximum at $y\approx 2$ and decreases
towards smaller and larger values of $y. $ Thus when $y=1,3$, it has values approximately
one half of  its maximum value.
Therefore, in order to crudely estimate the energy transfer one may assume that
only the contribution of eigenmodes having eigenfrequencies such  that $1 < y < 3$
need to be considered and that all these modes
 have $I_{2,-2}(y)=I_{2,-2}(y=2)$. The absolute values of the
eigenfrequencies are close to $\omega_{max}=2\Omega -2/ \eta,$ being
defined by the condition $y(\omega_{max})=2$. When $\eta \gg 1$ we can estimate the number of
participating  modes to be  $\propto 1/\eta $.  On the other hand, the typical frequency $\omega_{max}$
slightly increases when $\eta$ gets larger, tending asymptotically to $2\Omega .$  For the rotation rates considered
in this Paper the overlap integrals corresponding to $\omega_i\approx 2\Omega $ increase quite sharply with increasing
 $\omega_i$, see Figs. \ref{Qr1} and \ref{Qr1.5}. We have checked that the fact the $Q_i$ increase with  increasing
 $\eta$  and  $\omega_{max},$  approximately compensates for the  decrease in the number of modes giving
a sizable contribution to $\Delta E.$   Therefore, the sum of the contributions
of these modes behaves approximately as a constant
over  a range of values of $\eta.$  In this range $\Delta E\propto  \eta^{-2}$ and both $\Delta E_{I}$ and $\Delta L$
are negative. Note that this regime  persists  as long as the number of terms in the sum is larger than one, which
corresponds to $\eta < \eta_{max}\approx 2/\Delta \omega$, where $\Delta \omega$ is the distance between two neighbouring
eigenfrequencies having $\omega_i \approx \omega_{max}.$  When $\eta > \eta_{max},$  $\Delta E$ decreases faster than
$\eta^{-2}$. When $\Omega=0.42$ we find $\eta_{max}\approx 70$ for the Sun-like models and we have checked
that indeed this behaviour is observed in our results.


As discussed above the values of $\Delta E$ obtained  from the direct numerical approach for the non-rotating star are in
excellent agreement with the normal mode method, with only the values corresponding to $\eta =8$ deviating by about  20\%.
This deviation may be explained by the influence of numerical viscosity, which leads to relatively more dissipation
over  the long run times necessary when $\eta$  is large. Such runs become prohibitive for $\eta > 8.$
The case of $\Omega=0.11$ is quite similar to the non rotating case, with only one sizable deviation of
the order of 40\% associated with the retrograde encounter with $\eta=8.$
When $\Omega=0.21$ the deviations are  less than
than 25\%  for retrograde encounters and less than 30\% for prograde encounters with
$\eta \le 4\sqrt 2$. There is however, a large disagreement for the prograde encounter with $\eta=8$
for which the direct numerical approach
gives $\Delta E$ approximately 2.5 times larger than the normal mode method.
Convergence checks showed that this discrepancy was not due to lack of numerical
convergence of the direct numerical approach.

As seen from Fig. \ref{Fig18},
for $\Omega=0.21$ this $\eta $ is close to the value where $\Delta L=0$,  where the star
is in a state of pseudosynchronization
(see e.g. Papaloizou $\&$ Ivanov 2005, 2011, Ivanov $\&$ Papaloizou 2004, 2007
and references therein). In a similar problem of a tidal encounter of a  polytropic rotating star
 discussed in Papaloizou $\&$ Ivanov (2011),  an analogous
disagreement between the direct numerical and normal mode approaches was observed.

Close to the state
of pseudosynchronization
the  normal mode approach indicates that when $\eta$ is fixed and
$\Delta E$ and the absolute value of $\Delta L$ are considered to be functions of $\Omega,$ $\Delta E$ has a
deep minimum at $\Omega=\Omega_{ps},$ being the value for which pseudosynchronization occurs.  At  that point
$\Delta L=0.$ The differences between the two numerical approaches
may result in a shift in the location of this minimum, which because of its depth,
causes a large discrepancy when the methods are compared.

In addition, our normal mode approach does not take into account the contribution of inertial waves,
which can be excited in the convective regions and increase the amount of transferred energy as viewed in the rotating frame.
This effect would be most marked at pseudosynchronization.
One would expect that the inclusion of inertial modes would increase $\Delta E.$ To estimate the possible magnitude of the effect,
we note that Papaloizou \& Ivanov (2005) found that for a polytrope with $n=1.5,$
\begin {equation} \Delta E= 6.5\times 10^{-3}E_*/\eta^{6}.\label{PI2005}\end{equation}
Although only the convective envelope resembles such a polytrope, we use this estimate.
In fact there are two corrections, the first arising from the truncation of the envelope at $r\sim 0.7R_*,$ is expected
to increase $\Delta E$
by about an order of magnitude (Ogilvie 2013). The second, due to the fact that the envelope is on top of a more
centrally condensed model than the polytrope and so has a lower base density, is expected to decrease $\Delta E$ by a similar factor,
thus in order to make rough estimates, we simply assume these effects approximately cancel out.
Use of (\ref{PI2005}) for $\eta =8,$ gives $\Delta E \sim 2.5 \times 10^{-8}$ which is about five times the value indicated
in Fig. \ref{Fig17}. This indicates that inertial modes are likely to be  significant under conditions of pseudosynchronization
for $\eta=8.$ Similar estimates indicate that is is also the case for $\eta > 8.$

In support the above discussion,  we comment that
the  excitation of inertial waves
is seen  in our simulation of  the
prograde  encounter  of model 1b with $\eta=8$ and  $\Omega=0.21.$
To illustrate this,  contours of the product of the  Lagrangian displacement in
the $\phi$ direction and $\sqrt{\rho}$  are shown in an upper quadrant at a time after the encounter is over and the energy transfer has
been completed in Fig.\ref{Figsection}.
The square of this quantity is proportional to the kinetic energy density of the disturbance
associated with motion in the $\phi$ direction.
 The presence of rotationally modified $g$ modes
 is evident in the radiative core.   Inertial waves are seen in the convective envelope.
These show some reflections and graze the boundary with the radiative core as would be expected for critical latitude phenomena (see Papaloizou \& Ivanov 2010) .

When $\Omega=0.42$ the agreement
between two approaches is less good. The difference between $\Delta E$ is typically a factor of two
for the prograde encounters
and a factor of three for retrograde encounters at larger $\eta.$
The fact that such retrograde encounters give a larger disagreement can be
explained by the shift of the eigenfrequencies of the dominant  excited modes as viewed in the rotating frame.
These propagate against the sense of rotation of the star towards larger absolute values as $\eta$ increases.
In this case both the traditional and Cowling
approximations used in our normal mode approach become less appropriate.

In Fig. \ref{Fig18} we plot
 the amount of  angular momentum transferred, $\Delta L,$ as a function of $\eta.$
For  prograde encounters, in contrast to $\Delta E,$
$\Delta L$ changes sign at a value of $\eta,$ where the star rotates at the pseudosynchronization rate.
Accordingly, we plot the
absolute value of $\Delta L$ in this figure.
The form  of $|\Delta L,|$ for  prograde encounters is  non-monotonic,
having a deep minimum for  $\eta\equiv \eta_{1},$  where $\Omega=\Omega_{ps}$.
 When $\eta < \eta_{1},$ $\Delta L$ is positive
(ie. directed in the sense of stellar rotation), on the other hand  when $\eta > \eta_1,$
it is negative. In the case of retrograde encounters
$\Delta L$ is always positive (ie. directed in the sense of the orbital motion).
The behaviour of  the deviation between the direct numerical and normal mode approaches is  similar to
that found for  $\Delta E.$
We see  again a better agreement  for prograde encounters, with the exception of the encounter
having $\Omega=0.21$ and $\eta=8$, where the deviation is rather large. This may be explained as before.

Overall, our results indicate
quantitative agreement between the two approaches when the angular frequency is relatively small, say, $\Omega \le 0.2$
except for  rotation rates close to $\Omega_{ps}.$  For faster rotators the agreement is not so good with
the direct numerical approach  giving values of $\Delta E$ that are  a factor of $2-3$ smaller for retrograde encounters
and a factor of $2-3$
larger for prograde encounters as long as $\eta<8.$ These discrepancies probably arise from the neglect of inertial  waves
in the normal mode treatment as well as use of the traditional approximation and the neglect of self-gravity.
Excitation of inertial waves would cause $\Delta E$ to increase  near to pseudosynchronization  while the neglect
of self-gravity and the use of the traditional approximation become less appropriate for modes
excited at the high relative forcing frequencies that occurs for large retrograde stellar rotation.

In Fig. \ref{Fig19} we   plot the energy transfer in the inertial frame, $\Delta E_{I}$ related to $\Delta E$ and $\Delta L$ through
equation (\ref{eq1}).
As for the angular momentum transfer, it is negative for prograde encounters with $\eta > \eta_2$, where
$\eta_2\sim \eta_1$, and, therefore,  absolute values are  plotted.
The  energy exchanged for prograde
encounters has a sharp minimum at $\eta=\eta_2$, which moves  towards larger values of $\eta $ as the   magnitude
of $\Omega $ decreases.  Note that $\Delta E_I$ is always positive for  retrograde encounters. Let us recall that
solid, dashed, dot dashed and dotted curves apply to  $|\Omega|=0.42$, $0.21$, $0.11$ and $0$, respectively.
Together with these
results we also show the energy transfer calculated adopting
the 'perturbative' approach where it is assumed that the overlap integrals
are not modified by rotation and the eigenfrequencies can be calculated using  (\ref{eq0}).
The respective curves are represented
by dot dot dashed lines.  Symbols on these lines show different values of $|\Omega|$ with
circles, squares and diamonds corresponding to $|\Omega|=0.42$, $0.21$, $0.11$, respectively.
Remarkably, the perturbative approach
agrees quite well with the one based on the traditional approximation, especially for retrograde encounters, even for
the largest value of $|\Omega|=0.42$ adopted. In the case
of prograde encounters  there is quantitative agreement when $\eta < \eta_2$, and, accordingly, $\Delta E_I > 0$. Since
the perturbative approach does not require the calculation of the overlap integrals for every given value of $\Omega,$ the
evaluation of $\Delta E_{I}$ is simplified to a great extent. It  suffices to use the overlap integrals
obtained for non-rotating stars together with  the frequency splitting coefficients, $\beta_{r},$ given above for a number of stellar
models (see eg. Ivanov $\&$ Papaloizou 2011).

In Fig. \ref{Fig20} we plot  the characteristic timescale of evolution of the semimajor axis given   by equation (\ref{eq3})
as a function of $P_{orb}.$ The
line styles are as for  Fig. \ref{Fig17}. It is seen that for a given value of $P_{orb},$
retrograde encounters have smaller
values of $T_{ev}$ than the  corresponding  prograde encounter.  One can see, that for
fast rotators,  rotation has a significant influence on the strength of tidal encounters ( see also Lai 1997
and Ivanov $\&$ Papaloizou 2011). For example, when $|\Omega|=0.11,$\footnote{This  corresponds to a  rotation period of the star
of approximately  one day.} the binary system may significantly change its semimajor axis in  less than $10^9$yrs
for $P_{orb} < 2.7$days
and $< 4$days for  prograde and retrograde encounters, respectively.
Although stellar rotation significantly slows down in time
this effect may contribute to explaining   observed exoplanetary systems  containing Hot Jupiters with a significant mismatch
between directions of their orbital angular momentum  and the  rotation axis  of their  central stars.

In Fig. \ref{Fig21} we show the  tidal capture radius $R_{cap}$ for the rotating 1b model. As seen from this
Fig. the value of $R_{cap}$ is larger for retrograde encounters  as compared to
prograde encounters  as was  first noted  by Lai (1997).
However, this effect is prominent only when either,  the rotation rate is quite large,
 or the characteristic  relative velocity, $v_{*}$, is small.
It is instructive to compare the dependence of $R_{cap}$ on the rotation rate
found here with results of  Lai (1997),  bearing in mind, however, that
in that work,  a tidal encounter of a $n=1.5$ polytrope  having $M_{*}=0.4M_{\odot}$
and $R_{*}=0.5R_{\odot}$ with a point mass with $m_p=1.5M_{\odot}$  was considered.
 When $v_{rel}=2.5km/s$ we find  $R_{cap}\approx 2.7R_{*}$ and $4.2R_{*}$ for  prograde and retrograde encounters with
the rather large value,  $\Omega=0.42,$ On the other hand  Lai (1997)
gives  values of $R_{cap}\approx 4.2R_{*}$ and $6.4R_{*}$ for
the same  encounter,
but with  $\Omega=0.6$.
The ratio of the radii is approximately $0.65$ in both cases. Since the rotation rate of our model is smaller, the relative variation of
the  capture radius produced by changing from retrograde to prograde
rotation is somewhat larger in models of stars with realistic Sun-like structure as expected.

Finally, in Figs. \ref{Fig22}-\ref{Fig25} we plot the same quantities  as in Figs. \ref{Fig17}, \ref{Fig18}, \ref{Fig20} and
\ref{Fig21}, but for the rotating model 1.5a. The  absolute values of $\Omega$ are  $|\Omega|=0.5$, $0.25$ and $0.$
The results behave in a similar way  to those found for the  rotating Sun-like star,
 with the difference that the transfers of energy
and angular momentum for a given value of $\eta,$ are smaller for the more massive star,  as is found in  the non-rotating case.
The evolution timescales $T_{ev}$ are, however, larger for the Sun-like star due to its larger average density. For example,  model
1.5a with $\Omega=0.25$ gives an  evolution timescale of less  than $10^9$years for retrograde encounters,  when the orbital period after
circularisation $P_{orb} < 12$ days.
 Of course, as in the  previous case, in order to make realistic calculations,  one must take into
account the  evolution of the stellar structure as it affects the mean density of the star,
as well as the braking of the  stellar rotation with age.

\section{Conclusions and Discussion}

In this paper we have calculated the  energy and angular momentum transferred to
a number of Population I stellar models with different masses, ages and states of rotation through dynamical tides,
as a result of an encounter with a companion on a parabolic orbit. The results were used to estimate
the initial  evolution time scale of the semimajor axis of a highly
eccentric orbit.
Complementary methods based on calculation of  the  normal mode response and
a direct  numerical approach involving the solution of the encounter  problem as an initial value problem  were  used.
These showed  quantitative agreement for small and moderate rotation rates $|\Omega | <0.2$
as long as the distance of closest approach  was small enough that  the stellar angular velocity is  less than
the so-called pseudosynchronization frequency.

It was shown that when the energy and angular momentum transferred is  expressed in  natural units that factor out  the dependence on
stellar mass and radius so that encounters are characterised by the tidal parameter, $\eta,$
these quantities depend significantly
on the stellar structure.
The tidal transfers are significantly larger for models having sufficiently extended convective envelopes.
Thus, when $\eta=8,$ other things being equal, the non-rotating 1.5a model undergoes an energy transfer approximately $4$ times larger
than occurs for  models 1.5b and 1.5c
,which is explained by the presence of a  more extended convective envelope.
However, as the solar models 1a and 1b have more extended convective envelopes than  models  1.5b and 1.5c
which are older than model 1.5a,
the energy transferred to them is greater.
The effect increases in significance for larger values
of $\eta.$
Thus when $\eta=8,$ it is a factor of $15$ times larger.

Since the  models of more massive stars with $M_{*}=2M_{\odot}$ and
$5M_{\odot}$ that we considered
essentially do not have convective envelopes,  the energy transferred to them,  expressed in  natural units,
is rather small. Thus  model 5a has a  value of  $\Delta E$ ten times smaller than that  for  model 1b when $\eta=8.$
Stellar rotation was found to  play an important role, with dynamic tides being significantly amplified
for retrograde  encounters, and weakened
for prograde ones, see also Lai (1997) and Ivanov $\&$ Papaloizou (2011).

We studied  the effect of rotation using the direct numerical approach,
the normal mode approach adopting  the traditional approximation,
and also simply  treating the effects of rotation by  perturbation theory.
In the latter treatment, it was
assumed that the overlap integrals are not modified by rotation
but that  eigenfrequencies are shifted by an amount proportional
to the product of the
splitting coefficient $\beta_r$ and rotation frequency $\Omega,$
as expected from first order perturbation theory.

It was shown that the perturbative approach gives
results in quantitative agreement with the treatment based on  the normal mode approach with the traditional approximation,
even for fast rotators,  as long as  the energy transferred
in the inertial frame $\Delta E_{I}$ was positive,  being approximately equivalent
to the condition that the star rotated at
 less than the pseudosynchronization rate.

As implied by our discussion in section 4.5, this is also the condition for the forcing frequencies to be large enough
that the excitation of inertial modes is not expected to play an important role.
A condition for this to apply  for prograde encounters can be  approximately found by
requiring that the characteristic forcing frequency,
$\Omega_*/\eta,$ exceed $2\Omega.$ Making use of  equation (\ref{eq2}), we obtain the condition  as
\begin{equation}
P_{orb} < \frac{\Omega_*}{6.1\sqrt{\bar
\rho}\Omega}  ,\label{eq2a}
\end{equation}
Noting that for a typical rotation period of a T Tauri star of $6$ days and
solar parameters,  equation (\ref{eq2a}) gives $P_{orb} < 8.4$  days, the implication is that
the excitation of stellar inertial modes do not play a significant role
in the tidal capture of hot Jupiters into final  prograde orbits with periods of a few days.

This simplifies applications of the theory to particular systems since use of the the perturbative
approach requires only  the  calculation of the overlap integrals, and the splitting coefficients
for a given non rotating  stellar model.  These are provided
for all models considered in this paper.
In addition to the methods described above,
we also applied  the  purely analytic approach
developed in Paper 1 to the solar model 1b,
and showed that it gave results differing
from those obtained  numerically by  at most  40 per cent in  the range
$2 < \eta < 30$. It is important to stress again that the energy and
angular momentum transfers are significantly   larger for this model
as compared to  models of more massive stars and  a 'reference' model of $n=3$ polytrope.

	The stellar mean  density plays an important role in
applications to particular astrophysical systems. This is because
tides become relatively  more efficient for radially extended low-density objects.
In particular, timescale $T_{ev}$ for  the evolution of the  semimajor axis,
 considered as a function
of the orbital period after the period of circularisation $P_{orb},$ becomes smaller for
more rarefied evolved massive models regardless of the fact that other effects, such as possessing a smaller
convective envelope, act in the direction of making
tidal interaction less efficient  as discussed above.
Thus, $T_{ev}$ is less than $10^9$yrs for non-rotating models 1b
and 1.5c when $P_{orb} < 3.3$days and $ < 4.3$days, respectively.
For fast rotators this time can be significantly reduced for
retrograde encounters.
Thus, in the case of  a 1.5a model rotating with the angular frequency
$\Omega=0.25$ $T_{ev} < 10^{9}$yrs when $P_{orb} < 12$days.

When the theory of dynamic tides is
applied to particular astrophysical processes, such
as the tidal circularisation of exoplanet orbits starting with a high eccentricity
induced by gravitational scattering,
or the process
of tidal capture of stars in stellar clusters,
it is important
to calculate the overlap integrals and the coefficients $\beta_{r}$
for a grid of stellar models of different ages.
It is  also important to  understand the evolution of the stellar
structure, and rotation rate as a function of time.
The outcome of tidal
evolution may differ significantly
for stars with different masses and different rotational history.

As discussed in Paper 1 the overlap integrals are also important for discussing
the
tidal evolution of binaries with small orbital  eccentricity. In particular,
for forcing frequencies large enough that inertial modes
are not expected to be excited, they  fully determine the effect of tidal interactions in
the  regime of 'moderately large viscosity', see eg.
 Goodman $\&$ Dickson 1998 and Paper 1.
The results of Zahn (1977)  can only be  recovered
only when the overlap  integrals are
$\propto\, \omega^{17/6}.$ This dependence  approximately holds for Sun-like stars in the
limit of $\omega \rightarrow 0$. It  is  determined
by the functional form  of the square of the
Brunt-V$\ddot {\rm a}$is$\ddot{\rm a}$l$\ddot {\rm a}$
  frequency in the neighbourhood of the transition from radiative
to convective regions.  In particular the  $ \omega^{17/6}$   dependence of the overlap integrals
requires that, in the neighbourhood of the transition,  the square of the
 Brunt-V$\ddot {\rm a}$is$\ddot{\rm a}$l$\ddot {\rm a}$ frequency   is  a linear function of
the difference between a given radius and the radius of the transition.
This assumption may not be valid for a massive star, where the transition
from the  radiative envelope to  the convective core can be  extremely sharp.
In particular, the Zahn (1977)  theory does not apply to  binaries with
ultra-short periods $P_{orb}\sim 10\Omega_{*}^{-1}$, where the overlap
integrals decrease much faster with
$\omega,$ see the discussion of
Figs. \ref{Fig5}-\ref{Fig8} in the text. A theory appropriate for large  orbital periods must
consider the origin of, and take into account,  possible rapid  variations of the the
Brunt-V$\ddot{\rm a}$is$\ddot {\rm a}$l$\ddot {\rm a}$
frequency in the neighbourhood of the convective to radiative transition.
This is left for future work.

\section*{Acknowledgements}

We are grateful to I. W. Roxburgh for providing stellar models  and he and  G. I. Ogilvie for
fruitful discussions. We also thank S. V. Vorontsov for useful comments.

PBI and SVCh were supported in part by Federal programme
"Scientific personnel" contract 8422, by RFBR grant 11-02-00244-a,
by grant no. NSh 2915.2012.2 from the President of Russia and by
programme 22 of the Presidium of Russian Academy of Sciences.

Additionally, PBI was supported in part by the Dynasty Foundation
and thanks DAMTP, University of Cambridge for hospitality.

\label{lastpage}

\end{document}